\documentclass[12pt]{article}
\pdfoutput=1
\usepackage{graphicx}
\usepackage{subfigure}
\usepackage{amsfonts}
\usepackage{amssymb,amsmath}
\usepackage{hyperref}
\usepackage{comment}
\usepackage{multicol}
\usepackage{slashed}
\usepackage{multirow}
\usepackage[usenames,dvipsnames]{xcolor}
\definecolor{rosso}{cmyk}{0,1,1,0.3}
\definecolor{verde}{cmyk}{0.8,0,0.6,0.25}
\definecolor{bluc}{cmyk}{1,0.4,0,0.1}
\definecolor{blucc}{cmyk}{0.8,0.3,0,0}

\def\be{\begin{equation}}
\def\ee{\end{equation}}
\def\ba{\begin{eqnarray}}
\def\ea{\end{eqnarray}}

\def\GEV{\rm GeV}
\def\met{\slashed{E}_T}
\def\ntwo{\tilde{\chi}^0_2}
\def\nthree{\tilde{\chi}^0_3}
\def\none{\tilde{\chi}^0_1}
\def\cone{\tilde{\chi}^{\pm}_1}
\def\sl{\tilde{\ell}}

\def\gsim  {\hspace{0.3em}\raisebox{0.4ex}{$>$}\hspace{-0.75em}\raisebox{-.7ex}{$\sim$}\hspace{0.3em}}
\def\lsim  {\hspace{0.3em}\raisebox{0.4ex}{$<$}\hspace{-0.75em}\raisebox{-.7ex}{$\sim$}\hspace{0.3em}}

\makeatletter

\newcommand{\Rmnum}[1]{\expandafter\@slowromancap\romannumeral #1@}
\makeatother

\begin{document}
\begin{titlepage}
\begin{flushright}
KCL-PH-TH/2015-07 \\
LCTS/2015-03 \\
UT-15-06\\
\end{flushright}

\vskip 3cm

\begin{center}

{\Large\bf A closer look at a hint of SUSY at the 8 TeV LHC }

\vskip .5in

{
Philipp Grothaus$^{(a,}$\footnote{philipp.grothaus@kcl.ac.uk}$^)$, Seng Pei Liew$^{(b,}$\footnote{liew@hep-th.phys.s.u-tokyo.ac.jp}$^)$ and Kazuki Sakurai$^{(a,}$\footnote{kazuki.sakurai@kcl.ac.uk}$^)$ }

\vskip .3in

{\em
$^a$Department of Physics, King's College London, London WC2R 2LS, UK \vspace{0.2cm} \\
$^b$Department of Physics, University of Tokyo, Bunkyo-ku, Tokyo 113-0033, Japan \\
}

\begin{abstract}
A recent CMS analysis has reported the observation of an excess in the invariant mass distribution of the opposite-sign same-flavour lepton pair, which can be interpreted as a kinematic edge due to new physics. Using collider simulation tools, we recast relevant LHC search results reported by ATLAS and CMS collaborations in order to determine constraints on supersymmetric models that could produce the observed features. In particular, we focus on models involving cascade decays of light-flavour squarks and sbottoms. 
We find no favourable supersymmetry scenario within our exploration that could explain the origin of the excess when other LHC constraints are taken into account.
\end{abstract}

\end{center}
\end{titlepage}


\section{Introduction}
The search for supersymmetry (SUSY) as an extension of the Standard Model (SM) is one major target of the Large Hadron Collider (LHC) physics program. However, we have thus far found no definitive evidence of SUSY based on the first run of the LHC despite dedicated searches on many fronts. Even so, there is an analysis presented by the CMS collaboration that could be showing the first signs of SUSY~\cite{CMS-PAS-SUS-12-019,Khachatryan:2015lwa}. In the analysis, two leptons, jets and missing energy are looked for in the final states. It is found that there is an excess ($130^{+48}_{-49}$ events in the ``central" region) in the invariant mass distribution of the opposite-sign same-flavour (OSSF) lepton pair, corresponding to a significance of 2.6 $\sigma$.

The excess of the signal is fitted kinematically as a triangular-shape edge at $m_{\ell \ell}= 78.7 \pm 1.4~\GEV$. Such a kinematic edge is a characteristic signal of SUSY, where a SUSY particle undergoes a two-stage two-body decay. The kinematic edge formed by a pair of leptons can be interpreted as the cascade decay of a neutralino: $\ntwo \to \sl^{\pm}\ell^{\mp}\to \ell^{\pm}\ell^{\mp}\none$ (on-shell slepton decay)~\cite{Allanach:2000kt}. It is also possible to interpret the edge as a three-body decay signal of a neutralino, $\ntwo \to \ell^{\pm}\ell^{\mp}\none$, where the lepton pair is produced via an off-shell $Z$ (off-shell $Z$ decay). The shape of the edge would be more rounded as compared to the two-stage two-body decay, but as shown in the original CMS analysis, the three-body decay still provides a good fit. The direct  production of $\ntwo$ is too small to reproduce the dilepton excess, however its production can be boosted if coloured sparticles subsequently decay into $\ntwo$. The explanation 
of the dilepton excess in terms of coloured sparticles is consistent with the CMS analysis, since events with jet multiplicity are selected and counted.

In this work we perform a detailed study on the possibility of explaining the dilepton excess with several SUSY models
taking into account a comprehensive list of LHC constraints from a number of ATLAS and CMS direct SUSY searches.
In order to accurately estimate the LHC constraints and simulate many analyses systematically, 
we use the automated simulation tool {\tt Atom}~\cite{atom}.   
We take a bottom-up approach by considering simplified SUSY models with minimal content of particles at low energy to reproduce the excess optimally.
As will be discussed in the following sections, light-flavour squarks and sbottoms are potential candidates that satisfy these criteria.
Some of these models have already been studied in earlier works \cite{1409.3532, 1410.4998}.~\footnote{See~\cite{Allanach:2015ria} for a non-SUSY interpretation of the observed excess.}
Here, we will confront our simplified models with various direct SUSY search constraints such that their viability is tested in great detail.
We will show that the light-flavour squarks and sbottom models we consider in this paper are strongly constrained 
when providing a large enough contribution to the dilepton excess.

 Our paper is organised as follows. In the next section, we describe the selection criteria of the CMS dilepton analysis. In section~\ref{sc:susy}, we consider SUSY models that can possibly reproduce the required features of the dilepton edge. In section~\ref{sc:sim}, we describe the procedure of our simulation and analysis. We discuss our results and their interpretations in Section~\ref{sc:re}. Conclusions are drawn in Section~\ref{sc:con}.

\section{CMS dilepton analysis}

CMS reported an excess of events in the dilepton plus missing energy channel~\cite{CMS-PAS-SUS-12-019,Khachatryan:2015lwa} in the 8 TeV, 19.4 fb$^{-1}$ data.
The analysis requires an OSSF lepton pair with $p_T > 20$ GeV.
It also requires $\ge 2$ jets with $p_T > 40$ GeV and $\met > 150$ GeV or $\ge 3$ jets with $p_T > 40$ GeV and $\met > 100$ GeV.
The excess is observed in the central region where both leptons satisfy $|\eta_{\rm  lep}| < 1.4$.
It exhibits an $edge$ in the dilepton invariant mass distribution around $m_{ \ell \ell} = 78$ GeV.
The counting experiment in the $m_{ \ell \ell} \in [20,70]$ GeV region 
shows an excess of $\sim 130$ events over the Standard Model expectation, which corresponds to a standard deviation of 2.6 $\sigma$.

\section{SUSY interpretations of the dilepton edge}
\label{sc:susy}

In this paper we consider simplified SUSY models that capture the essence needed for explaining the observed dilepton excess. Generalizations of SUSY models given here are straightforward. 

It is known that the OSSF dilepton pair in the decay of the second lightest neutralino $\ntwo$ via on-shell slepton
and off-shell $Z$ exhibit an edge-like shape at
\ba
m_{\rm edge} = m_{\ntwo} \sqrt{ \left(1 - \frac{m^2_{\sl}}{m^2_{\ntwo}}\right) \left( 1 - \frac{m^2_{\none}}{m^2_{\sl}} \right) }&:&~~~\ntwo \to \sl^{\pm} \ell^\mp \to \ell^\pm \ell^\mp \none,
\label{eq:edge1}
\\
m_{\rm edge} = m_{\ntwo} - m_{\none}&:&~~~\ntwo \to \ell^\pm \ell^\mp \none,
\label{eq:edge2}
\ea
respectively, in the $m_{\ell \ell}$ distribution.
In order to obtain a large enough production cross section to fit the excess and to have $\ge 2$ high $p_T$ jets required in the event selection,
we consider production of coloured SUSY particles, which may subsequently decay into $\ntwo$.   
In this paper we consider two scenarios: light-flavour squark and sbottom production.

\subsection{Squark scenarios}

In the squark scenario, we consider the production of pairs of light-flavour squarks. 
This scenario assumes the first two generations of squarks (both left and right-handed) to be mass degenerate
and within the reach of the LHC, whilst the third generation squarks and gluinos are decoupled.
We also assume that the second lightest neutralino is mostly Wino-like or an admixture of Wino and Higgsinos
and the lightest neutralino is mostly Bino-like.
In this setup the lighter chargino, $\cone$, is naturally introduced as a $SU(2)_L$ partner of the $\ntwo$
and their masses have to be close to each other.   
Since the right-handed squarks do not couple to the Wino and only very weakly couple to the Higgsinos, they decay predominantly into a quark and the $\none$,
whereas the left-handed squarks can decay to either $\cone$, $\ntwo$ or $\none$ depending on the Wino-Higgsino mixing in the $\cone$ and $\ntwo$.

We consider two models according to the $\cone$ and $\ntwo$ decay modes.  
The first model is the {\it on-shell slepton} model, where we assume the right-handed selectron and smuon in the low energy spectrum
so that the $\ntwo$ decays predominantly into an OSSF lepton pair and the $\none$ via the on-shell slepton.
We decouple the left-handed slepton doublets, ($\tilde \nu_L, \tilde \ell_L$), to maximise the signal rate, otherwise
the $\ntwo$ could also decay into a pair of neutrinos and the $\none$ via the on-shell $\tilde \nu_L$.\footnote{In our setup the $\ntwo$ decays predominantly into muon pairs through the Higgsino component of the $\ntwo$.  
}

Any models that lead to multi-lepton final states are severely constrained by the multi-lepton plus missing energy searches~\cite{CMS-PAS-SUS-13-002}.
To avoid these constraints a large branching ratio of the $\tilde q_L \to q \none$ mode is necessary in this model.
We assume 

\be
BR( \tilde q_L \to q + \cone / \ntwo / \none ) = 10 / 5 / 85 \, \%.
\ee
This can be achieved if $\cone$ and $\ntwo$ have large Higgsino components because the squarks couple to the Higgsinos with small Yukawa couplings. We will see in section~\ref{sc:re_sq} that
our conclusion is not sensitive to variations of the branching ratios.

In the squark with on-shell slepton model we then have the following cascade decays
\ba
\tilde q_L \to q \ntwo \to q \ell^\pm \sl^\mp \to q \ell^\pm \ell^\mp \none  &:&  ~~~5\,\%,   \nonumber
\\
\tilde q_L \to q \cone  \to q \nu \sl^\pm \to q \nu \ell^\pm \none  &:&  ~~~10\,\%,  \nonumber
\\
\tilde q_L \to q \none &:& ~~~85\,\%, \nonumber 
\\
\tilde q_R \to q \none &:& ~~~100\,\%.  \nonumber 
\ea
If one of the pair produced squarks undergoes the first decay chain, the final state may contain
an OSSF dilepton plus two energetic jets, and such events can contribute to the CMS dilepton excess.

\begin{figure}[t!]
\centering
\includegraphics[page=1,width=12cm]{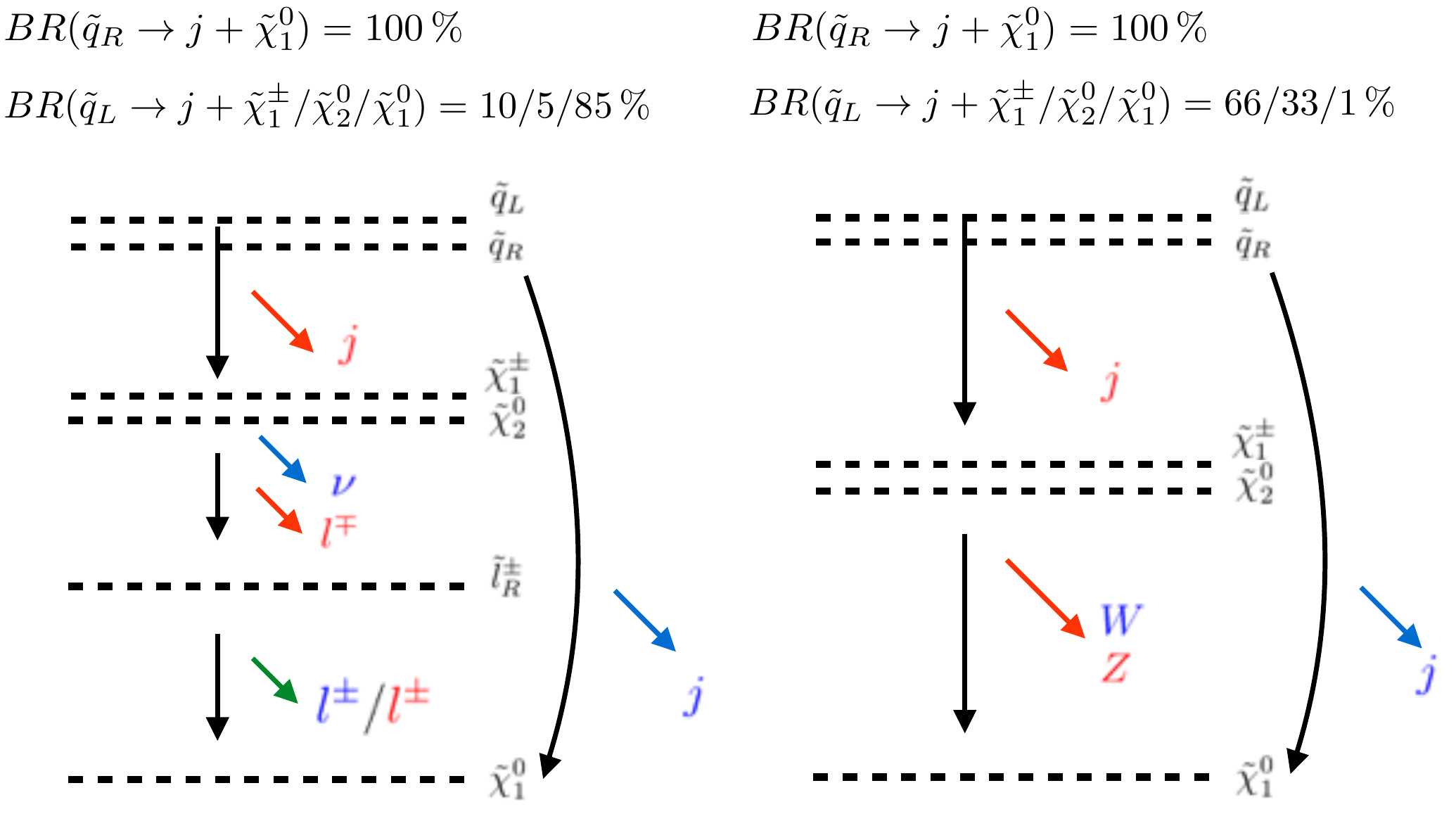}
 \caption{Decay chains of squark scenarios ({\it left}: on-shell slepton model, {\it right}: off-shell $Z$ model).} 
 \label{fig:squark_model}
\end{figure}

The second model we consider in this paper is the {\it off-shell $Z$} model, where the $\ntwo$ decays  via the off-shell $Z$ into an OSSF dilepton pair and the $\none$.
Unlike in the on-shell slepton model we here need a large branching ratio of $\tilde q_L \to q \ntwo$ such that the small leptonic branching ratio of the off-shell $Z$ into electrons and muons (about 6 \%)  is compensated. 
We assume
\ba
BR( \tilde q_L \to q + \cone / \ntwo / \none ) = 66 / 33 / 1 \, \%,
\ea
which can be realised by assuming $\cone$ and $\ntwo$ are Wino-like. 
In the squark with off-shell $Z$ model we have the following decay chains
\ba
\tilde q_L \to q \ntwo \to q f \bar f \none~({\rm via}~Z^*)  &:&~~~ 33\,\%,   \nonumber
\\
\tilde q_L \to q \cone \to q f \bar f^\prime \none~({\rm via}~W^*)  &:&~~~66\,\%,  \nonumber
\\
\tilde q_L \to q \none  &:&~~~ 1\,\%, \nonumber 
\\
\tilde q_R \to q \none &:&~~~ 100\,\%.  \nonumber 
\ea
The signal events can be obtained if one of the pair produced squarks undergoes the first decay chain 
and the $\ntwo$ decays via the $Z^*$ into the dilepton pair and the $\none$.
A schematic picture of the squark scenarios is shown in Fig.~\ref{fig:squark_model}.

\subsection{Sbottom scenarios}
\label{sec:sbottom_scenario}
 
Another way of interpreting the CMS dilepton excess is to assume that the observed dileptons in the excessive events come from 
cascade decays of bottom squarks.
Unlike in the squark scenario, the decay mode to charginos, $\tilde b_1 \to t \cone$, is kinematically forbidden if $m_{\tilde b_1} < m_t + m_{\cone}$.
We consider this case because the decay mode to charginos is more constrained due to emergence of top quarks.  
Similarly to the squark scenario we consider {\it on-shell slepton} and {\it off-shell $Z$} models according to the $\ntwo$ decay mode.

In the on-shell slepton scenario the $\ntwo$ may decay either via a right-handed charged slepton or a left-handed charged slepton and sneutrino. We will treat these two cases separately in our analysis.

If the mediating slepton is right-handed, $\ntwo$ predominantly decays into two charged leptons and $\none$, and the events tend to have more than two leptons in the final state.
Such models are severely constrained by the multi-lepton plus missing energy searches as we have previously discussed.
To avoid these constraints, we assume  70\,\% of sbottoms decay into a bottom quark and a $\none$ and the rest of sbottoms decay into a bottom quark and a $\ntwo$.
This situation can be achieved if  $\ntwo$ is Wino-like and  $\tilde b_1$ has a large component of $\tilde b_R$.
We have the following decay chains in the sbottom with on-shell slepton model.
\ba
\tilde b_1 \to b \ntwo \to b \ell^\pm \sl^\mp \to b \ell^\pm \ell^\mp \none  &:&  ~~~30\,\%,   \nonumber
\\
\tilde b_1 \to b \none &:& ~~~70\,\%.  \nonumber 
\ea

\begin{figure}[t!]
\centering
 \includegraphics[page=1,width=12.5cm]{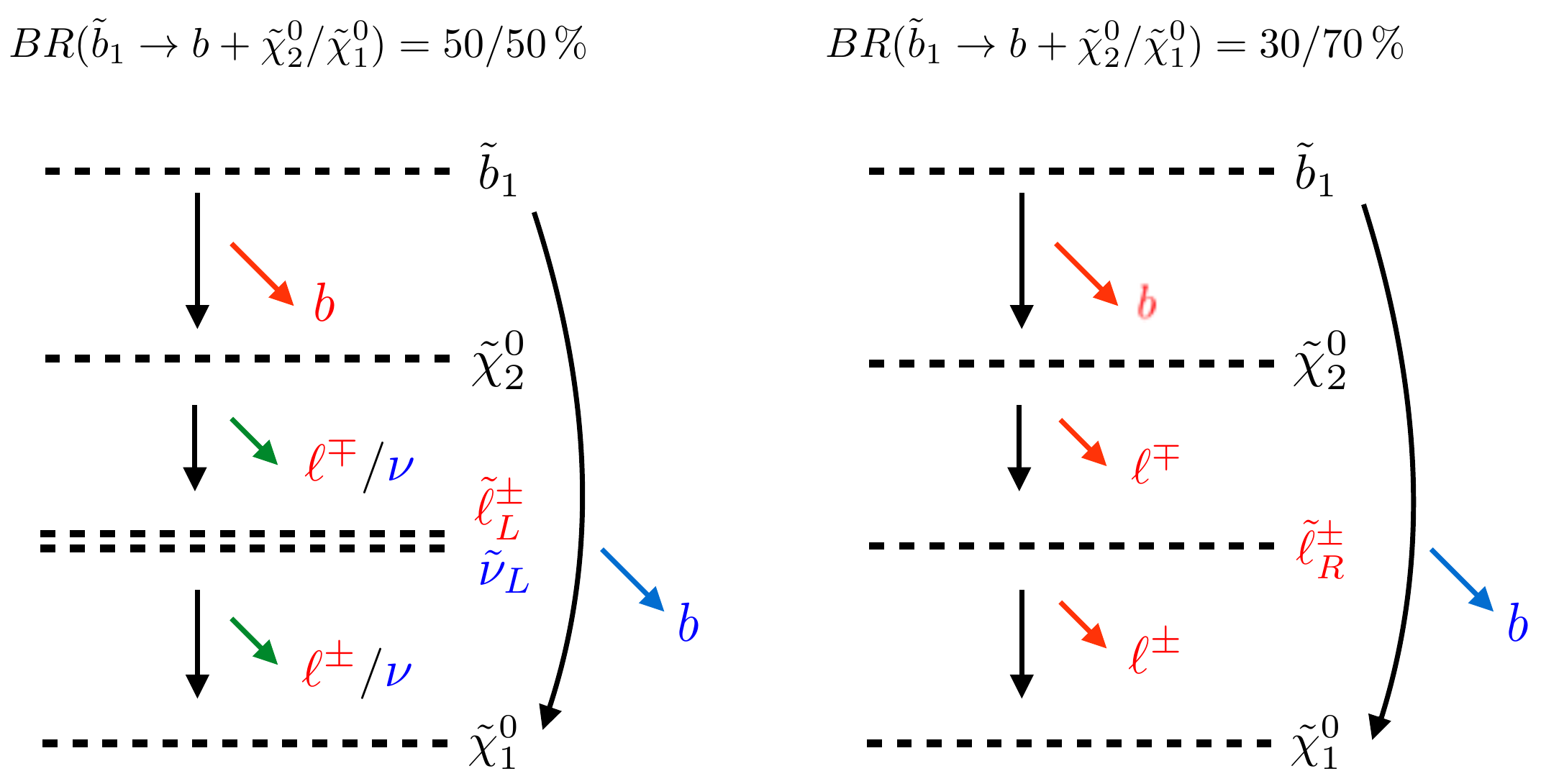}
  \caption{Decay chains of on-shell slepton mediated sbottom scenarios ({\it left}: left-handed slepton model, {\it right}: right-handed slepton model).} 
 \label{fig:sbottom_slep}
\end{figure}

In the case where the mediating slepton is left-handed, sneutrinos are introduced as SU(2) partners of charged sleptons.
We assume sneutrinos and charged sleptons are mass degenerate and 
$\ntwo$ decays democratically into charged sleptons and sneutrinos. 
\ba
\tilde b_1 \to b \ntwo \to b \ell^\pm \sl^\mp \to b \ell^\pm \ell^\mp \none  &:&  ~~~25\,\%,   \nonumber
\\
\tilde b_1 \to b \ntwo \to b \nu {\tilde \nu} \to b \nu \nu \none  &:&  ~~~25\,\%,   \nonumber
\\
\tilde b_1 \to b \none &:&  ~~~50\,\%.   \nonumber
\ea
The schematic picture of these cases is shown in Fig.~\ref{fig:sbottom_slep}.

Let us discuss the off-shell $Z$ model for the sbottom scenario. Analogous to the squark with off-shell $Z$ model, we need sbottoms to have a sizeable decay branching ratio to $\ntwo$ in order to have large enough dilepton event rates.
One way to realise this situation is to have a Higgsino-like $\ntwo$, a mostly right-handed $\tilde b_1$ and to assume a large sbottom-bottom-Higgsino coupling due to a large ${\rm tan} \beta$. 
It is shown in~\cite{1410.4998} that for ${\rm tan} \beta=50$, $m_{\tilde b_1} \simeq 330~\GEV$ and a Higgsino mass parameter $\mu\simeq 290$ around 44\% of sbottoms decay to the roughly mass-degenerate $\ntwo$ and $\tilde \chi_3^0$.  
This model point predicts about 1-$\sigma$ less events than the central fit without being excluded. 
In order to explore in more detail the parameter region that could possibly contribute to the excess, 
we expand the study of this scenario by varying the parameters $M_1$, $\mu$ and $m_{\tilde b_1}$, while fixing ${\rm tan} \beta=50$. 
The mass spectrum and particle decay branching ratio of this simplified model are calculated using {\tt SPheno}~\cite{Porod2003,Porod2011}. 

\begin{figure}[t!]
\centering
 \includegraphics[page=1,width=10.5cm]{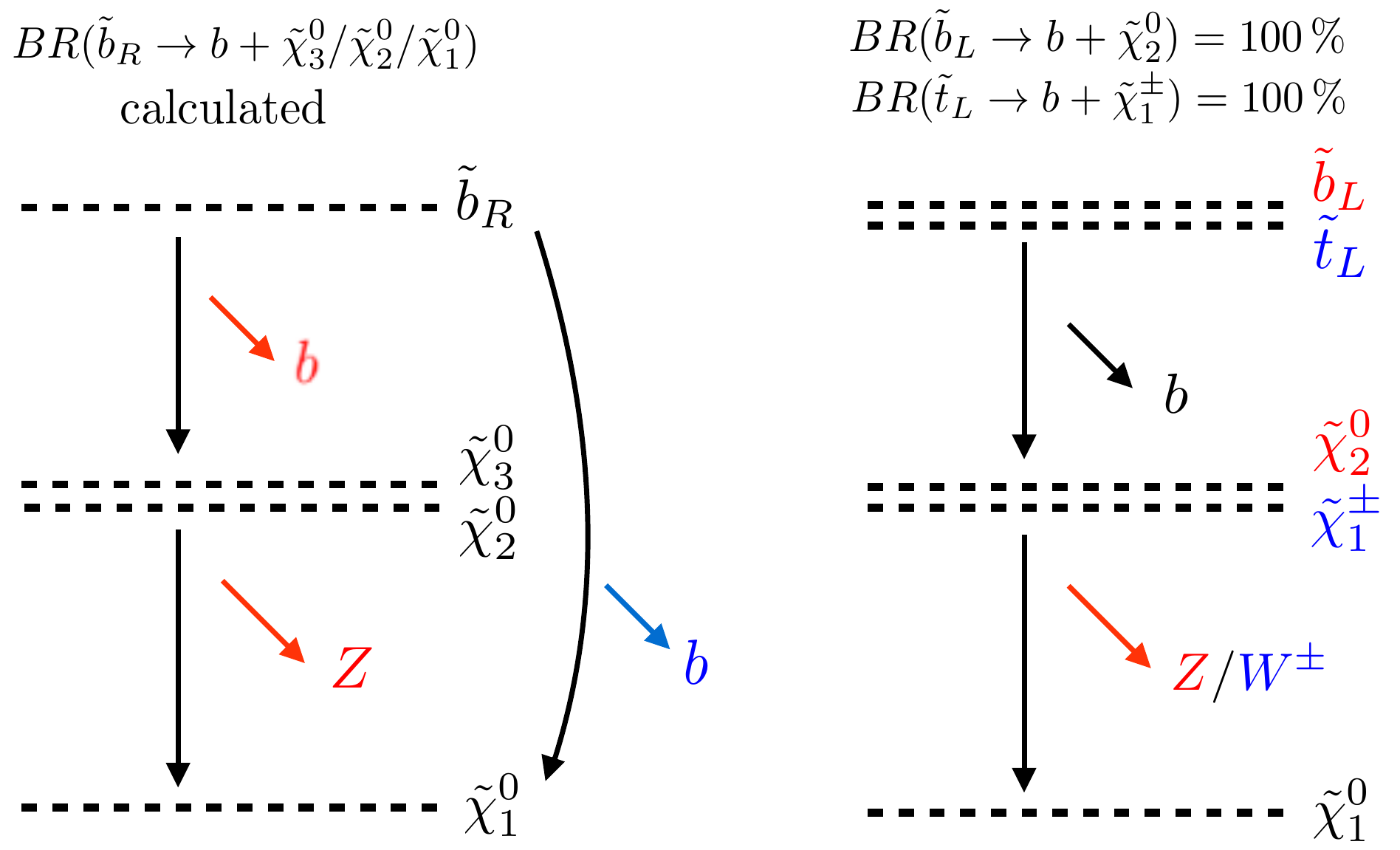}
  \caption{Decay chains of off-shell $Z$ mediated sbottom scenarios ({\it left}: right-handed sbottom model, {\it right}: left-handed sbottom model).} 
 \label{fig:sbottom_Z}
\end{figure}

Alternatively one can obtain a large branching ratio of sbottom decaying to $\ntwo$ by assuming $\tilde b_1$ is left-handed and $\ntwo$ is Wino-like. 
Due to $SU(2)$ gauge invariance a left-handed top squark, $\tilde t_1$, is necessarily included in the low energy spectrum. 
For simplicity, we assume $m_{\tilde b_1}=m_{\tilde t_1}$. We consider the following decay chains for the left-handed sbottom with off-shell $Z$ model.
\ba
\tilde b_1 \to q \ntwo \to q f \bar f \none~({\rm via}~Z^*)  &:&~~~ 100\,\%,   \nonumber
\\
\tilde t_1 \to q \cone \to q f \bar f^\prime \none~({\rm via}~W^*)  &:&~~~100\,\%.  \nonumber
\ea
A schematic picture of the off-shell $Z$ sbottom scenarios is shown in Fig.~\ref{fig:sbottom_Z}.

The squark and sbottom scenarios are the priorities of this work, but let us also touch on the possibilities of explaining the dilepton excess with the remaining coloured sparticles in SUSY, namely gluino and stop. Gluinos can decay into $\ntwo$ via an intermediate squark, not much different from the squark or sbottom scenario other than a larger jet multiplicity. For stop, its decay into a top quark would lead to an extra lepton that plays no role at explaining the dilepton excess. It is not clear how gluino or stop could explain the dilepton excess without inducing additional jet or leptonic constraints, and hence we are not going to study these scenarios further in this work.

\section{The simulation setup}
\label{sc:sim}

In this section we describe our procedure to calculate the contribution to the CMS dilepton excess 
and the constraints from other ATLAS and CMS SUSY searches.

The production cross section, $\sigma_{\rm prod}$, for light-flavour squarks is calculated using {\tt Prospino 2}~\cite{Beenakker1996} with the gluino mass set to 3.5 TeV. For the sbottom cross section we use results from the LHC SUSY Cross Section Working Group based on~\cite{Kramer2012}.
We create SLHA files of our simplified models for the event generation and pass them to {\tt Pythia 6.4}~\cite{Sjostrand2006} to generate a total number of $10 \cdot \sigma_{\rm prod } \cdot \mathcal{L}$, with maximal $5 \cdot 10^5$, events, where $\mathcal{L} = 19.4$ fb$^{-1}$ is the integrated luminosity at the CMS dilepton analysis.
We then run {\tt Atom}~\cite{atom} on the generated HepMC event files
to estimate the efficiencies, $\epsilon$, of the signal regions defined in all the ATLAS and CMS analyses that will be used in this work.
The application examples and validation of {\tt Atom} is found in \cite{Papucci:2011wy, Papucci:2014rja, Kim:2014eva}.
We have implemented the CMS dilepton analysis in {\tt Atom}
and validated it using the cut-flow tables given by the CMS collaboration based on the $\tilde b_1 \to b \ntwo \to b \ell^+ \ell^- \none$ simplified model. 
The comparison in the number of expected signal events calculated by {\tt Atom} and CMS is shown in Appendix \ref{ap:validation}.
We also cross-checked some of the analyses with another simulation tool {\tt CheckMATE}~\cite{1312.2591}.

From the obtained cross section and efficiency, the SUSY contribution to the CMS dilepton excess is calculated as
$N_{\ell \ell} = \sigma_{\rm eff} \cdot {\cal L}$, where the effective cross section, $\sigma_{\rm eff}$,
is defined as the cross section after the event selection: $\sigma_{\rm eff} = \epsilon \cdot \sigma_{\rm prod}$.
For the other ATLAS and CMS analyses the 95\,\% CL upper limit on $\sigma_{\rm eff}$, $\sigma_{\rm UL}$, 
is reported for each signal region by the collaborations.
We define a useful measure for exclusion by $R = \sigma_{\rm eff}/\sigma_{\rm UL}$.
If $R > 1$ is found for one of the signal regions, the model is likely to be excluded, although one needs to combine all the signal regions statistically 
to draw a definite conclusion.
However, we do not attempt to combine these signal regions because there are non-trivial correlations among them which originate from the uncertainties on $e.g.$ the jet energy scale, the lepton efficiency and luminosity,
and it is not possible for us to combine the signal regions correctly.
Instead, in the next section we will look at the exclusion measure $R$ individually to understand 
which signal regions are sensitive to the model points.

\begin{table}[t] 
\centering 
\begin{tabular}{|l|l|l|l|}
\hline
 channel & search for   & arXiv or CONF-ID & refs 
\\
\hline
{$2{\rm -}6 j + 0 \ell +  \met$}        & $\tilde{q},\tilde{g}$   		                            &ATLAS-CONF-2013-047
& \cite{ATLAS-CONF-2013-047} \\ 
&&1405.7875&\cite{1405.7875}
 \\
$2 b + 0 \ell + \met$ 	            & $\tilde{t},\tilde{b}$ 	                                    &1308.2631
&\cite{1308.2631}\\
\hline
{$4j + 1 \ell +\met$}             & $\tilde{t}$                				     &ATLAS-CONF-2013-037
& \cite{ATLAS-CONF-2013-037} \\
$\geq 2 j + \geq 1 \ell + \met$            & $\tilde{q},\tilde{g}~(1~{\rm or}~2  \ell)$                                   &ATLAS-CONF-2013-062
& \cite{ATLAS-CONF-2013-062} \\
\hline
{$2j + 2 \ell + \met$}          & dilepton edge 	                                            & CMS-PAS-SUS-12-019
& \cite{CMS-PAS-SUS-12-019,Khachatryan:2015lwa}\\
$2j + \ell^{\pm} \ell^{\pm} + \met$ & $\tilde{q},\tilde{g},\tilde{t},\tilde{b}~({\rm SS\ lepton})$ &ATLAS-CONF-2013-007
& \cite{ATLAS-CONF-2013-007} \\
$2j + 2 \ell + \met$            & $\tilde{t} (2 \ell)$                                             &ATLAS-CONF-2013-048
& \cite{ATLAS-CONF-2013-048} \\
&&1403.4853&\cite{1403.4853}\\
\hline
{$2,3 \ell + \met$}             & $\tilde{\chi}^{\pm},\tilde{\chi}^0,\tilde{ \ell }$                &1404.2500
& \cite{1404.2500} \\
&&1405.7570 &\cite{1405.7570} \\
$ 3 \ell + \met$                & $\tilde{\chi}^{\pm},\tilde{\chi}^0$                          &1402.7029
&\cite{1402.7029} \\
$\geq 3 \ell + \met$         & $\tilde{\chi}^{\pm},\tilde{\chi}^0$                          &CMS-PAS-SUS-13-002
&\cite{CMS-PAS-SUS-13-002} \\

\hline
\end{tabular} 
\caption{LHC searches used in this paper to test the viability of the simplified models.}
\label{tab:search} 
\end{table} 

In Table \ref{tab:search} we list the ATLAS and CMS analyses we consider in this work.
We include the multi-jet \cite{ATLAS-CONF-2013-047, 1405.7875} and di-$b$ jet \cite{1308.2631} analyses,
jets plus single \cite{ATLAS-CONF-2013-037} or two lepton \cite{CMS-PAS-SUS-12-019,Khachatryan:2015lwa, ATLAS-CONF-2013-007} 
(including same-sign (SS) dilepton \cite{ATLAS-CONF-2013-007}) analyses \cite{ATLAS-CONF-2013-062}
and multi-lepton analyses \cite{1404.2500, 1405.7570, CMS-PAS-SUS-13-002, 1402.7029}.
In the next section we investigate whether the SUSY models can fit the CMS dilepton excess taking the constraints from these analyses into account.

\section{Results}
\label{sc:re}
\subsection{Squark scenarios}
\label{sc:re_sq}
\begin{figure}
\centering
\includegraphics[width=7.9cm]{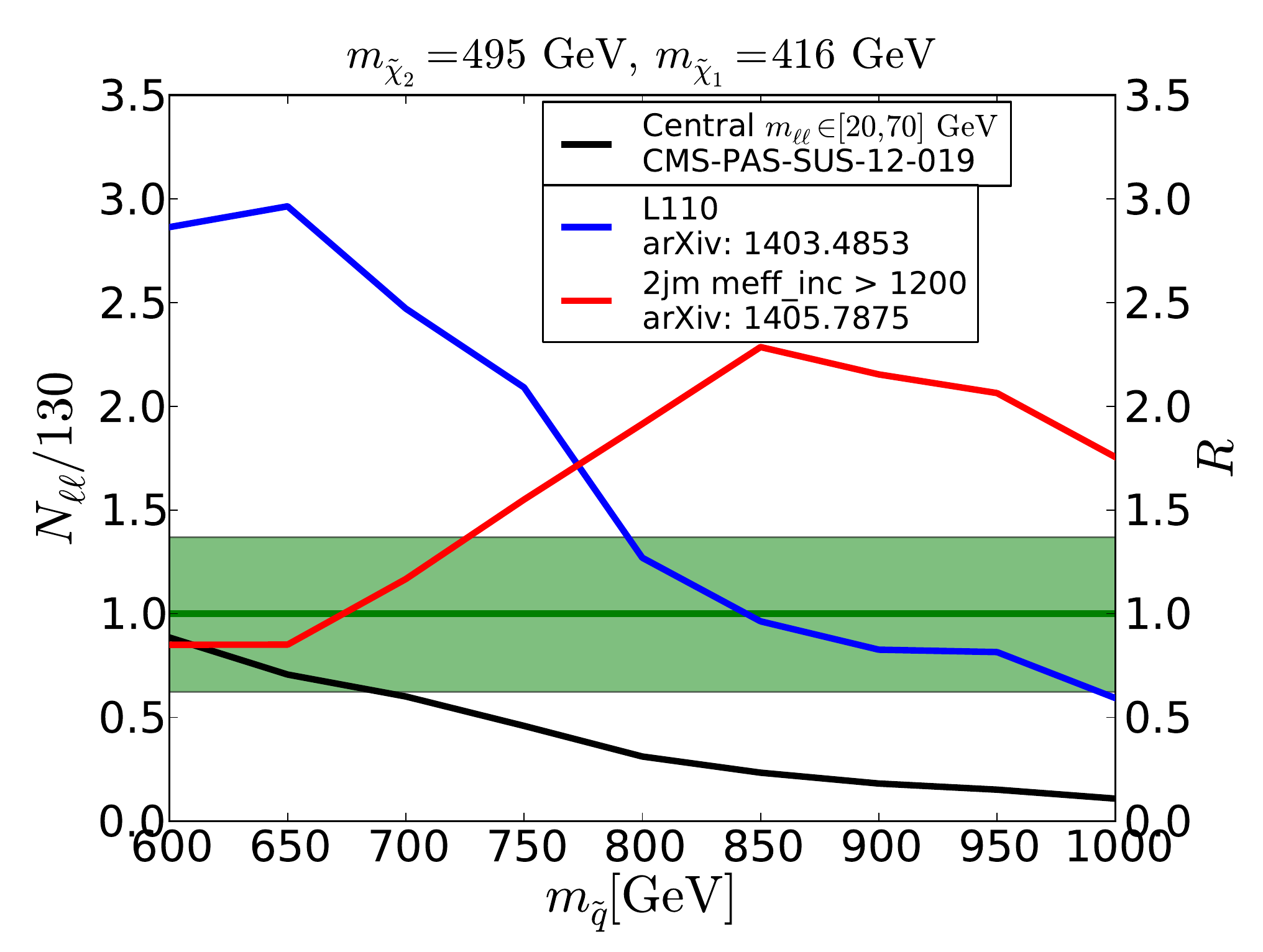}
\includegraphics[width=7.9cm]{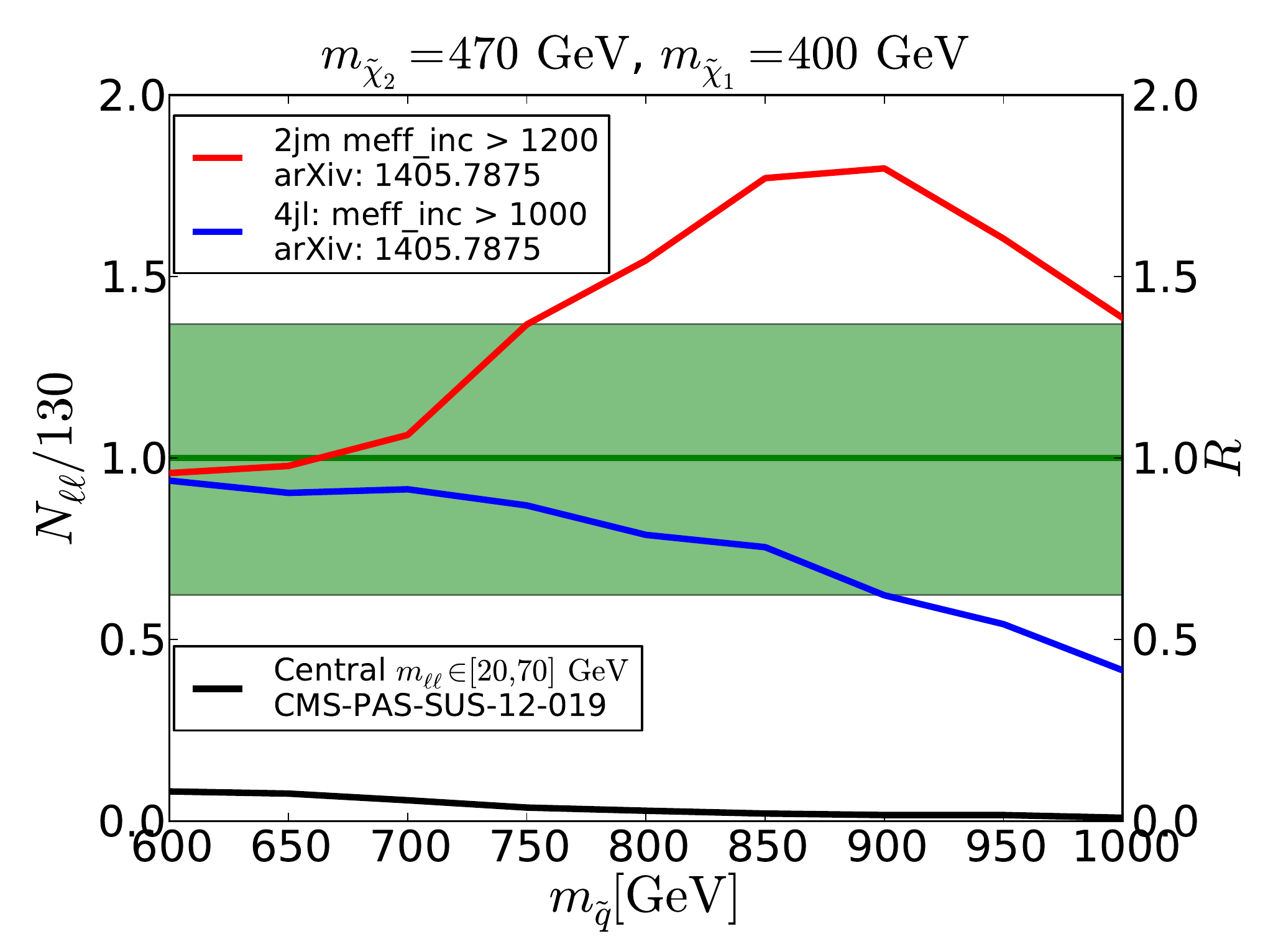}
 \caption{Signal rate and $R$-values for the squark models. The left panel presents the intermediate slepton, the right panel the off-shell $Z$ scenario.
 } 
 \label{fig:squark_results}
\end{figure}

In Fig.~\ref{fig:squark_results} we show the results of our numerical calculation for the squark scenario.
In the plots the black curves represent the SUSY contribution, $N_{\ell \ell}$, normalised by the best fit value 130.
The green bands correspond to the 1\,$\sigma$ region of the fit.
In the same plots we show also the exclusion measure, $R$, for a few signal regions that are particularly sensitive to the models.
The region where any $R$ is greater than 1 is strongly disfavoured.

In the left panel of Fig.~\ref{fig:squark_results} we show $N_{\ell \ell}/130$ and $R$ as functions of $m_{\tilde q}$ for the squark with on-shell slepton model.
We fix $m_{\ntwo} = 495$ GeV and $m_{\none} = 416$ GeV.
For these masses, there are no constraints from the chargino-neutralino direct searches.
The right-handed slepton mass is fixed at 450 GeV such that $m_{\rm edge}$ in Eq.~(\ref{eq:edge1}) is 78 GeV, which is the optimal value for the CMS dilepton excess. 

As can be seen, this model can fit the excess only in the region where $m_{\tilde q} \lsim 650$ GeV. 
However, this region is
strongly disfavoured by the L110 signal region (shown in the blue curve) in the ATLAS stop search \cite{1403.4853}. 
This signal region requires the same final state ($2 j + 2 \ell + \met$) as the CMS dilepton analysis,
in particular an OS lepton pair with $p_T > 25$ GeV and at least two jets with $p_T > 20$ GeV.
The condition $m_{T2} > 110$ GeV is also imposed, which is very effective to reduce the $t \bar t$ and $WW + {\rm jets}$ backgrounds.    
One can see that the sensitivity of this signal region decreases as the $m_{\tilde q}$ increases due to the reduction of the production cross section.
Nevertheless, the signal rate in the dilepton excess also decreases in the same way since these analyses employ similar event selection.  
Consequently there is no region in the plot where the SUSY events can fit the dilepton excess avoiding the exclusion from the other searches.
This conclusion is robust against our assumption on the branching ratios, $Br(\tilde q_L \to q + \cone/\ntwo/\none) = 10/5/85$\,\%,
since the L110 signal region constrains the same channel as in the CMS dilepton analysis. 

One can also see that in the $m_{\tilde q} > 680$ GeV region, the 2jm signal region in the ATLAS multi-jet search \cite{1405.7875}
becomes sensitive and rules out the model points.
This signal region is characterised by the requirement of 
at least two jets with $p_T > 130$ and 60 GeV
and a moderately large effective mass, $m_{\rm eff} \equiv \sum_i |p_{Ti}^{j40}| + \met > 1200$ GeV, where $p_{Ti}^{j40}$ is the $i$-th 
high $p_T$ jet with $p_T > 40$ GeV.
The events with an electron or muon with $> 10$ GeV are rejected in this analysis.   
The 2jm signal region targets the $\tilde q \tilde q \to q \none q \none$ topology, 
which is indeed the dominant event topology in this model since $Br(\tilde q_R \to q \none) = 100$\,\% and $Br(\tilde q_L \to q \none) = 85$\,\%.\footnote{
We note that in \cite{1405.7875} ATLAS does not exclude the region where $m_{\none} = 416$ GeV in the squark-neutralino simplified model.
We, on the other hand, exclude this neutralino mass for a certain range of the squark mass (see Fig.~\ref{fig:squark_results} (left)).
This is because our squark production cross section is larger than the ATLAS's value
since we set the gluino mass at 3.5 TeV in which the contribution from the gluino exchange diagram is still sizeable. 
}
Due to the harsh cut on the $m_{\rm eff}$, the 2jm signal region is sensitive to 
the models with large mass gaps between $\tilde q$ and $\none$.
This is the reason why the sensitivity increases as $m_{\tilde q}$ increases 
until the point ($m_{\tilde q} \simeq 850$ GeV) at which a rapid degradation of the squark production cross section 
finally turns the sensitivity down.

In the right panel of Fig.~\ref{fig:squark_results}, we plot the $N_{\ell \ell}/130$ and $R$
as functions of $m_{\tilde q}$ for the squark with off-shell $Z$ model,
where we fix $m_{\ntwo} = 478$ GeV and $m_{\none} = 400$ GeV
so that $m_{\rm edge}$ in Eq.~(\ref{eq:edge2}) is 78 GeV.
One can see that the SUSY contribution is too small to account for the dilepton excess,
whilst this region is severely constrained by the 2jm and 4jl signal regions in the ATLAS multi-jet search \cite{1405.7875}.
Compared to the on-shell slepton model, the rate of an OSSF lepton from a squark cascade decay is small:
$Br(\tilde q_L \to q \ntwo) \cdot Br(\ntwo \to Z^* \none) \cdot Br( Z^* \to \ell^+ \ell^-) \simeq 0.33 \cdot 1 \cdot 0.06 \simeq 2 \,\%$,
though we took a maximal value $33\,\%$ for $Br(\tilde q_L \to q \ntwo)$ assuming
$\ntwo$ and $\cone$ to be Wino-like.
Instead, $\ntwo$ and $\cone$ have large branching ratio to hadronic modes via $Z^*$ and $W^*$
which makes the off-shell $Z$ model more prone to be excluded by the ATLAS multi-jet search \cite{1405.7875}
compared to the on-shell slepton model due to the lepton veto cut in the analysis. 
The 2jm signal region constrains mostly $\tilde q_R \tilde q_R \to q \none q \none$ topology 
and the sensitivity peaks around $m_{\tilde q} \simeq 900$ GeV with $m_{\none} = 400$ GeV,
similarly to the on-shell slepton model.
On the other hand, the 4jm signal region requires 
at least 4 jets ($p_T > 130, 60, 60, 60$ GeV) and
looks at the jets not only from the squark decay, $\tilde q \to j \tilde \chi$ ($\tilde \chi = \none, \ntwo$ or $\cone$),
but also from hadronic $\cone$ and $\ntwo$ decays and initial state radiation.
Due to the milder cut $m_{\rm eff} > 1000$ GeV, the sensitivity peaks at a much lower squark mass.

We conclude that for the squark models it is very difficult to fit the observed CMS dilepton excess
if the ATLAS stop search \cite{1403.4853} and the ATLAS multi-jet search \cite{1405.7875} 
are both considered.

\subsection{Sbottom scenarios}
\label{sec:res_sbottom}
\begin{figure}[t]
\centering
\includegraphics[width=7.5cm]{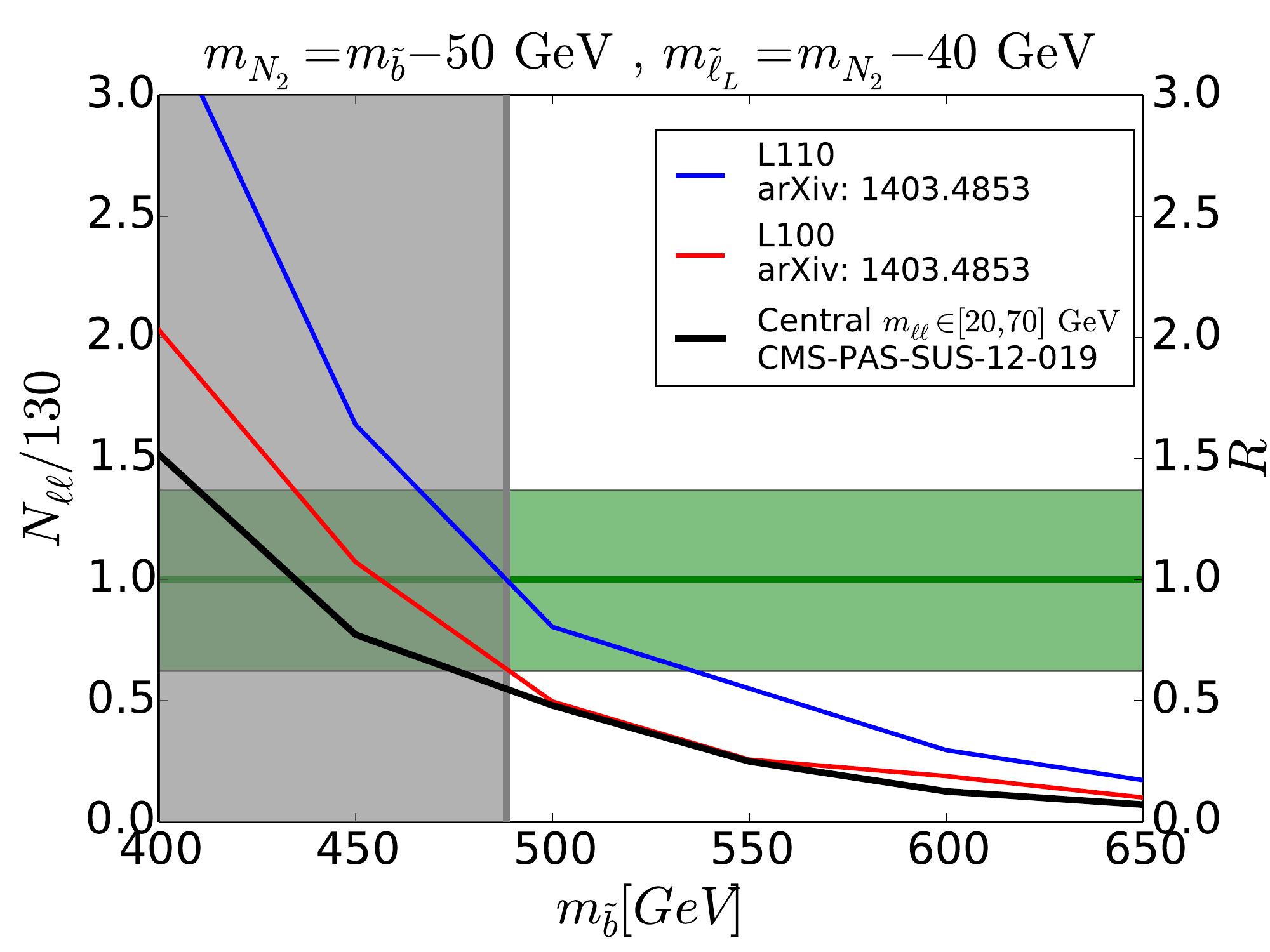}
\includegraphics[width=7.5cm]{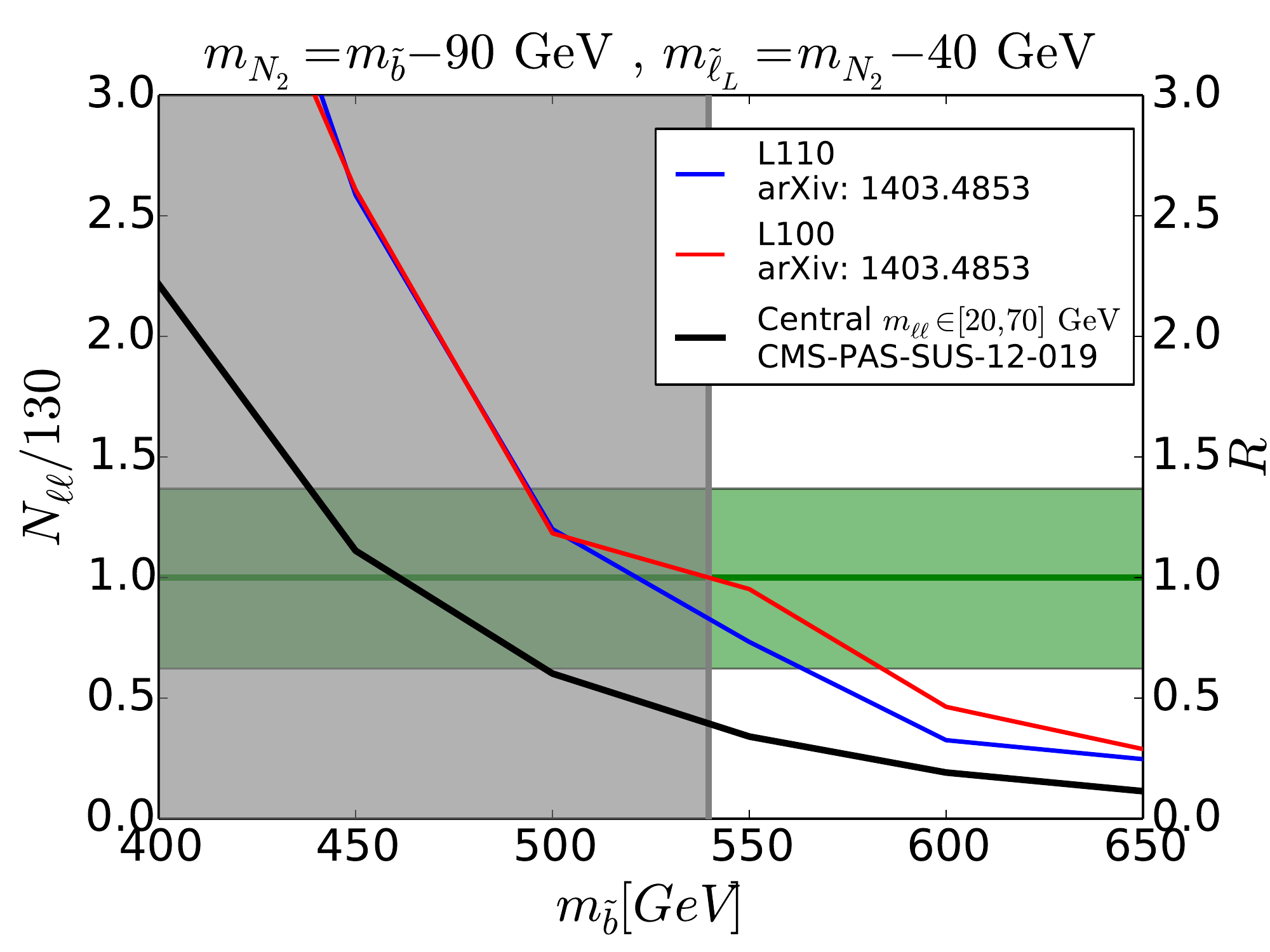}
\includegraphics[width=7.5cm]{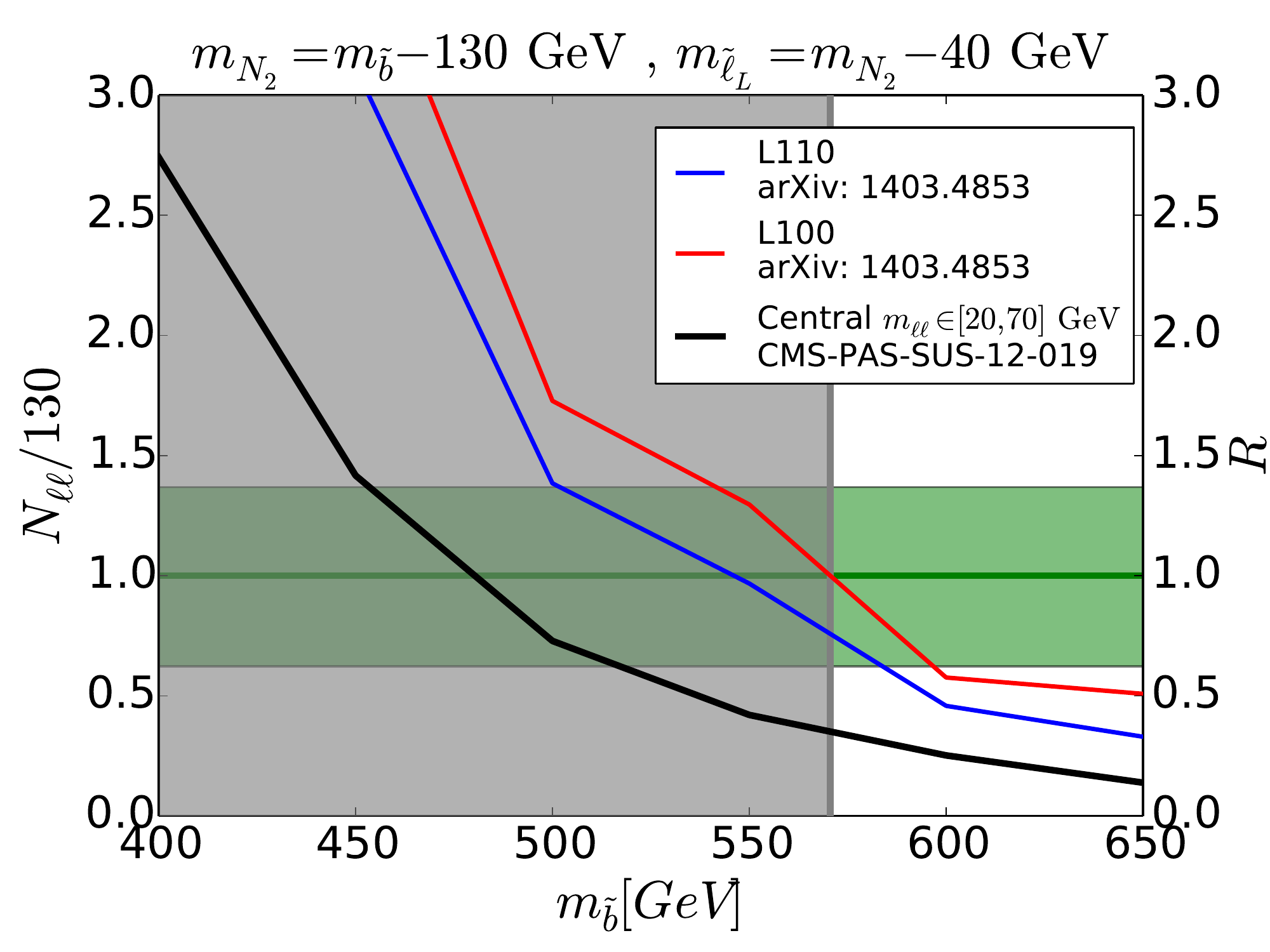}
\includegraphics[width=7.5cm]{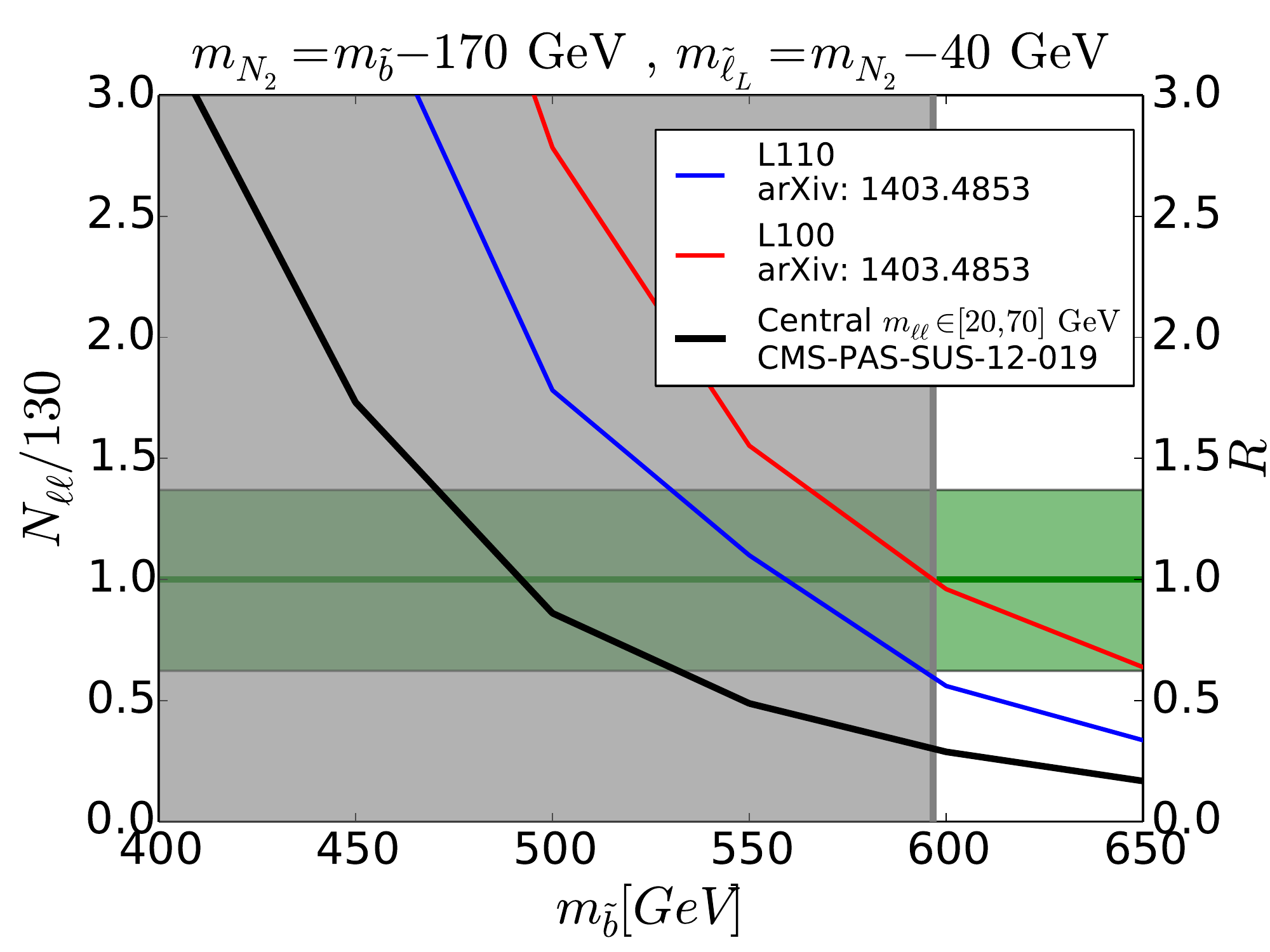}
 \caption{Signal rate and $R$-values for the on-shell left-handed slepton mediated sbottom models.} 
 \label{fig:sbottom_Lslepton}
\end{figure}

In this section we present the results for the sbottom scenarios, starting with the on-shell left-handed slepton model.
In Fig.~\ref{fig:sbottom_Lslepton} we show $N_{\ell \ell}/130$ and $R$ of the most constraining signal regions
as functions of $m_{\tilde b_1}$. 
As discussed previously, we assume $m_{\ntwo} < m_{\tilde b_1} - m_t$ to avoid tops in the decay chains that would lead to more stringent constraints. 
Within this condition we examine four different mass gaps: $\Delta m \equiv m_{\tilde{b}_1} - m_{\ntwo} =$~50, 90, 130 and 170 GeV. 
The left-handed slepton mass is fixed at $m_{\tilde{\ell}_L} = m_{\ntwo} - 40$ GeV and $m_{\none}$ is set for each combination of $m_{\ntwo}$ and $m_{\tilde{\ell_L}}$ such that $m_{\rm edge}$ in Eq.~(\ref{eq:edge1}) is 78 GeV. 
The intermediate slepton can either be a sneutrino or a charged slepton
and the branching ratio of $\ntwo$ into these two states is assumed to be equal.
Therefore, only half of the produced $\ntwo$ decay into an OSSF dilepton and a $\none$. 

In Fig.~\ref{fig:sbottom_Lslepton} we see that a good fit can be obtained for sbottom masses between 420 and 520 GeV, depending on $\Delta m$. 
However, these model points are strongly disfavoured by the L100 and L110 signal regions of the ATLAS stop search~\cite{1403.4853}. 
The event selection in the L100 signal region is very similar to the L110 signal region which we briefly described in the previous subsection. The difference is that in the L100 signal region the lepton $p_T$ requirement is raised to $(p_T^{\ell 1}, p_T^{\ell 2}) > (100, 50)$ GeV and $m_{T2} > 100$ GeV is imposed. As the lepton $p_T$ requirement is raised with respect to L110, L100 is especially sensitive to larger mass gaps.

\begin{figure}[t]
\centering
\includegraphics[width=7.5cm]{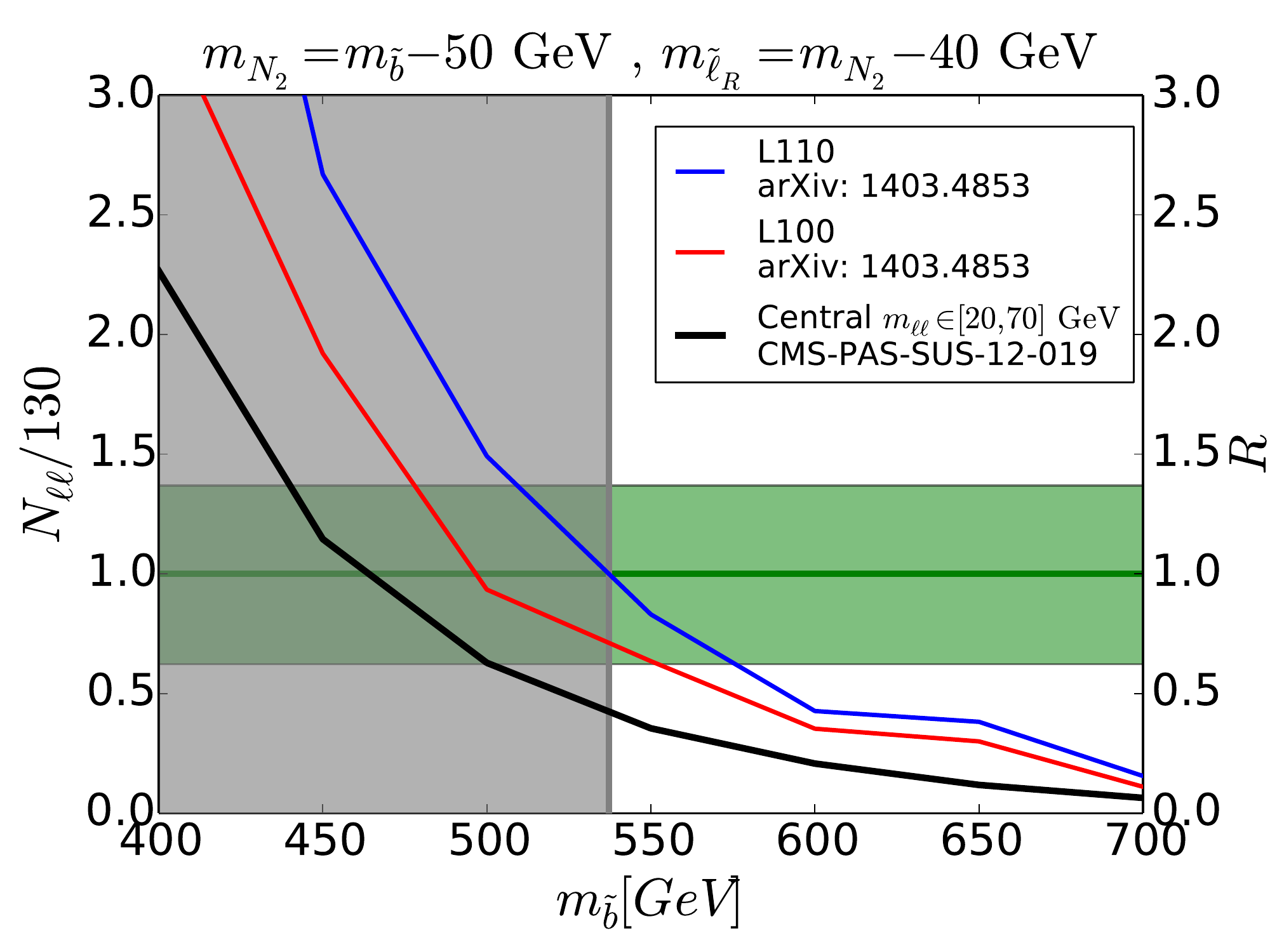}
\includegraphics[width=7.5cm]{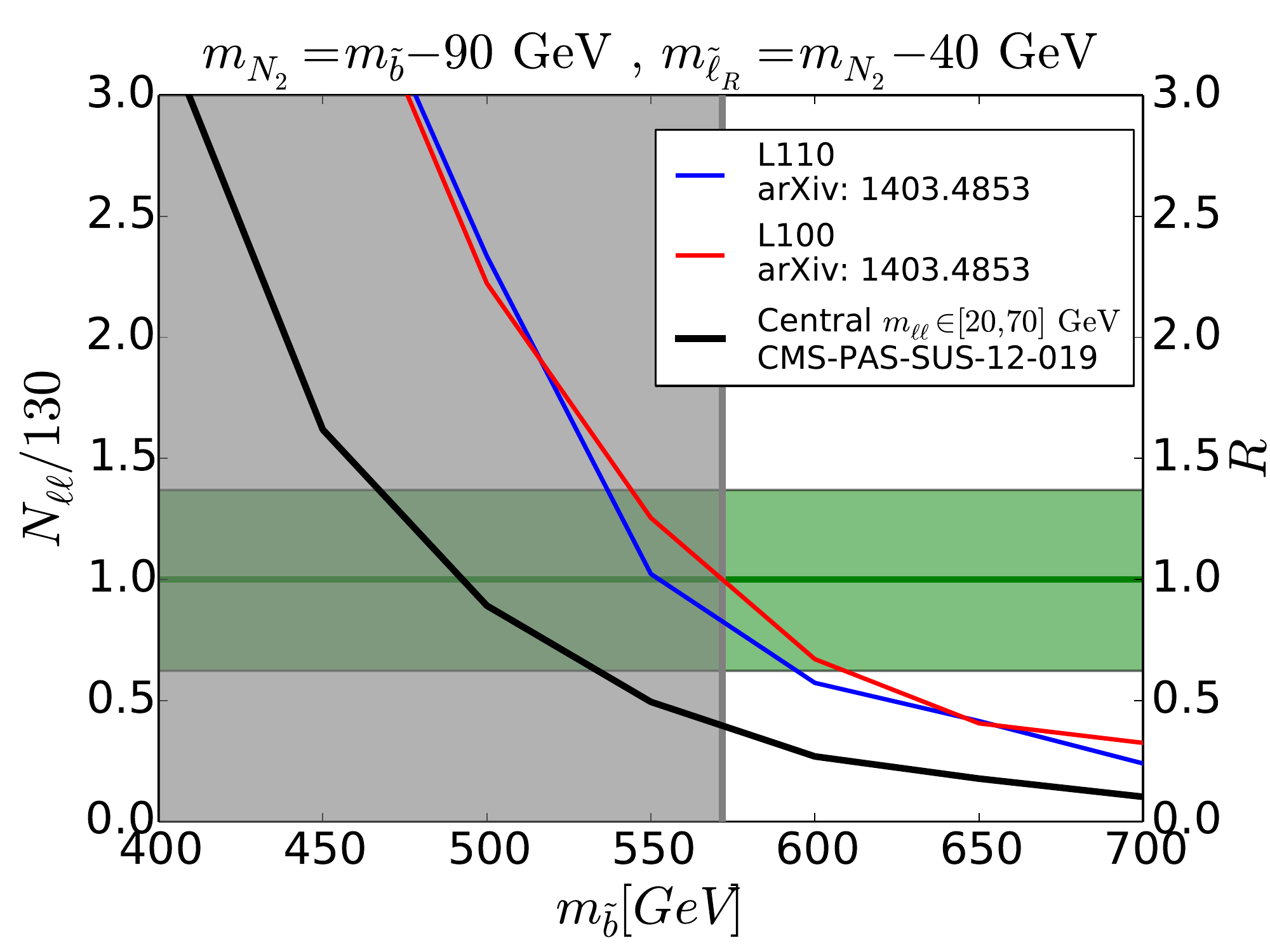}
\includegraphics[width=7.5cm]{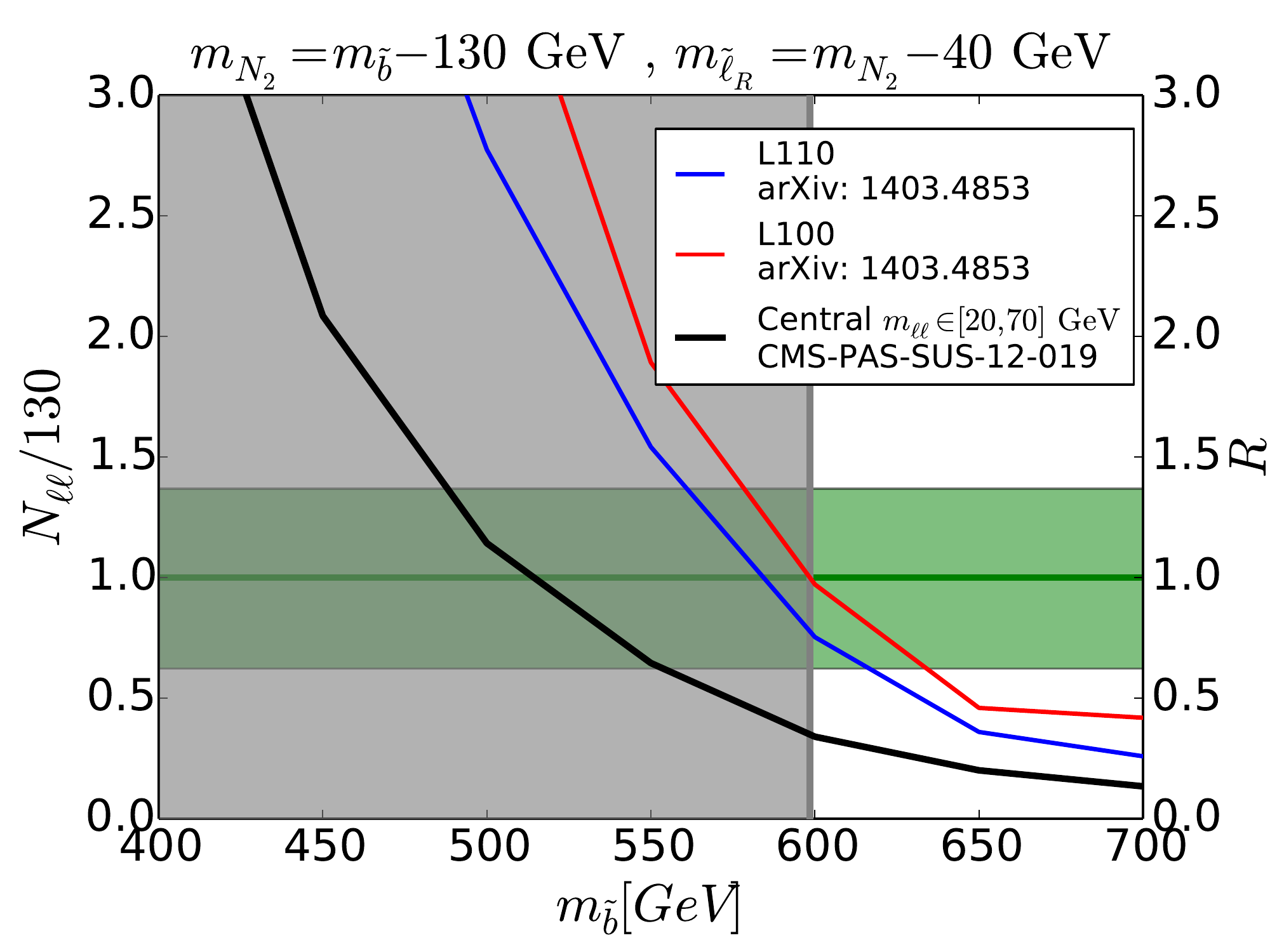}
\includegraphics[width=7.5cm]{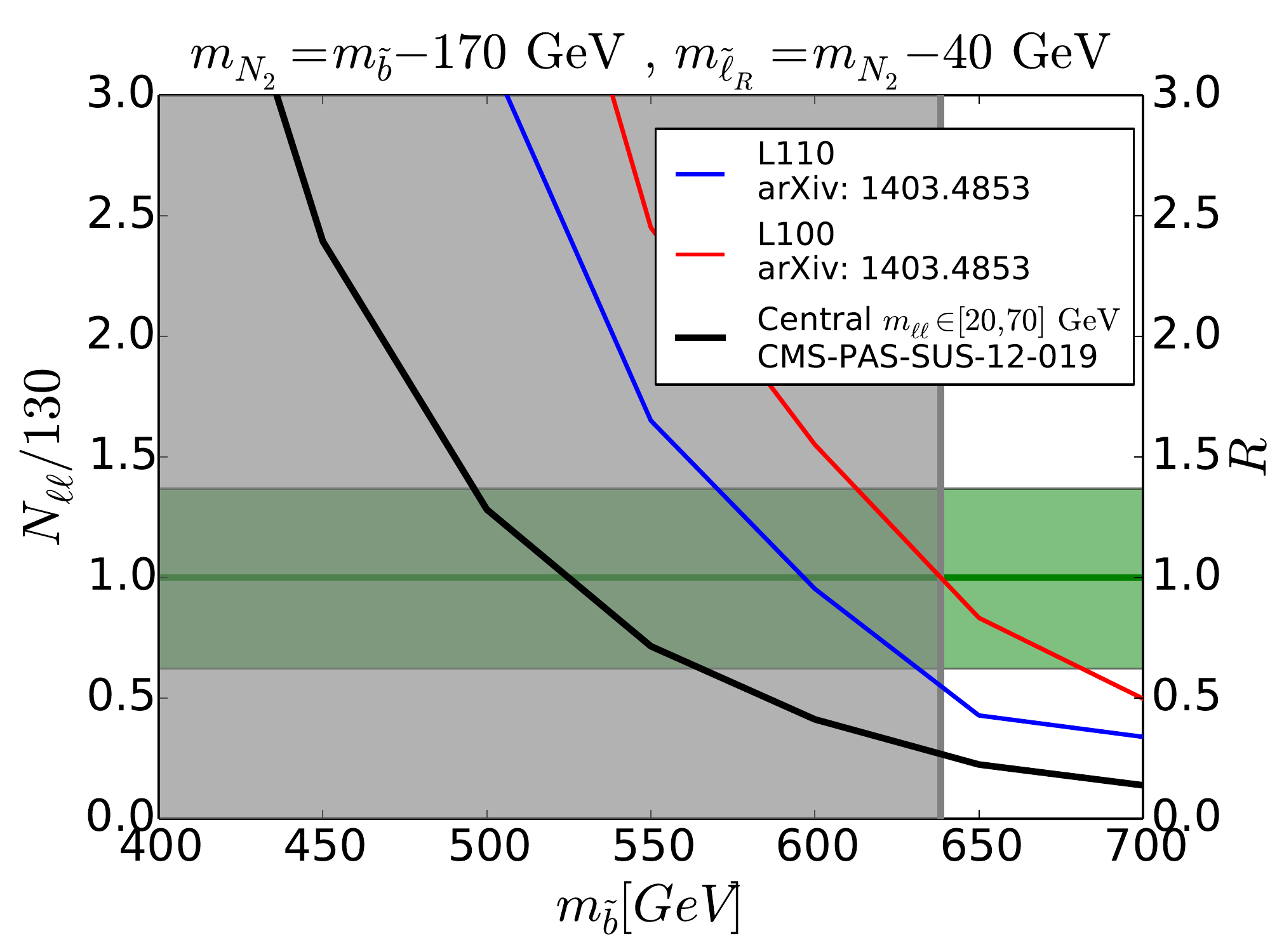}
 \caption{Signal rate and $R$-values for the on-shell right-handed slepton mediated sbottom models.} 
 \label{fig:sbottom_Rslepton}
\end{figure}

In Fig.~\ref{fig:sbottom_Rslepton} we show the contribution to the excess and the constraints from other searches
in the on-shell right-handed slepton model for the four different $\Delta m$, similarly to Fig.~\ref{fig:sbottom_Lslepton}.
In this scenario $\ntwo$ decays into an OSSF dilepton and a $\none$ with the branching ratio of 100\,\%.
As can be seen, the results are similar to the left-handed slepton case and
the region where the model gives a good fit is strongly disfavoured by the L110 and L100 signal regions of the ATLAS stop search.
The similarity of the results amongst the left- and right-handed slepton scenarios 
can be understood because L110 and L100 
constrain the same final state ($2j + 2 \ell + \met$) as that is targeted in the CMS dilepton analysis
and the kinematics of the dilepton events are similar between these scenarios.
We conclude that it is difficult to attribute the CMS dilepton excess to 
the sbottom with on-shell slepton models
if the constraint from the ATLAS stop search \cite{1403.4853} is taken into account.

We now turn to the sbottom with off-shell $Z$ models.
The first model we investigate is the right-handed sbottom model where $\ntwo$ and $\nthree$ are mostly Higgsino-like and $\none$ is mostly Bino-like.
In this model the masses of three lightest neutralinos are calculated from the parameters, $\mu, M_1$ and $\tan\beta$,
fixing $M_2$ at 3.5 TeV.
Since we assume $\mu > M_1$, we have $m_{\nthree} \sim m_{\ntwo} \sim \mu$ and $m_{\none} \sim M_1$
and both $\ntwo$ and $\nthree$ can contribute to the excess 
through their decays into an off-shell $Z$ boson and a $\none$. 
The decay rate of the sbottom into the Higgsino states is dictated by the
sbottom-bottom-Higgsino coupling which is proportional to $\tan\beta$.
In order to have a large signal rate, we take $\tan\beta = 50$ in our numerical scan. 
We again examine four different mass gaps 
$\Delta m =$ 50, 90, 130, and 170 GeV between the sbottom and $\ntwo$. 
To this end we vary $\mu$ such that $\ntwo$ takes the desired mass set by $\Delta m$.
$M_1$ is chosen such that $m_{\ntwo} - m_{\none} = 70$~GeV.
The mass of the lightest sbottom is calculated from given parameters fixing the left-handed third generation squarks mass, $m_{\tilde Q_3}$,  at 1.5 TeV.
A table with paramter values for each model point can be found in the Appendix~\ref{ap:pmssm}.

\begin{figure}[t]
\centering
\includegraphics[width=7.5cm]{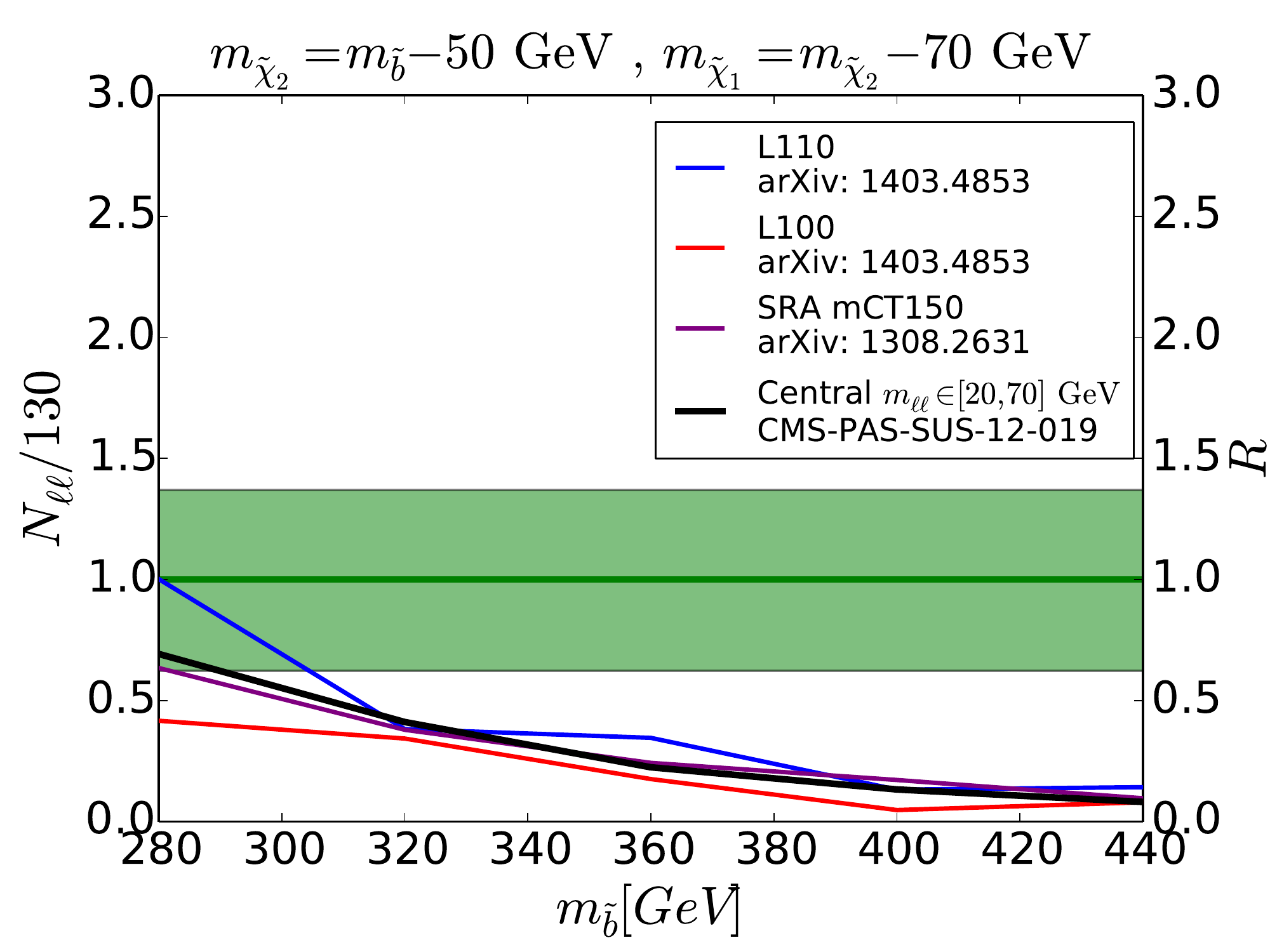}
\includegraphics[width=7.5cm]{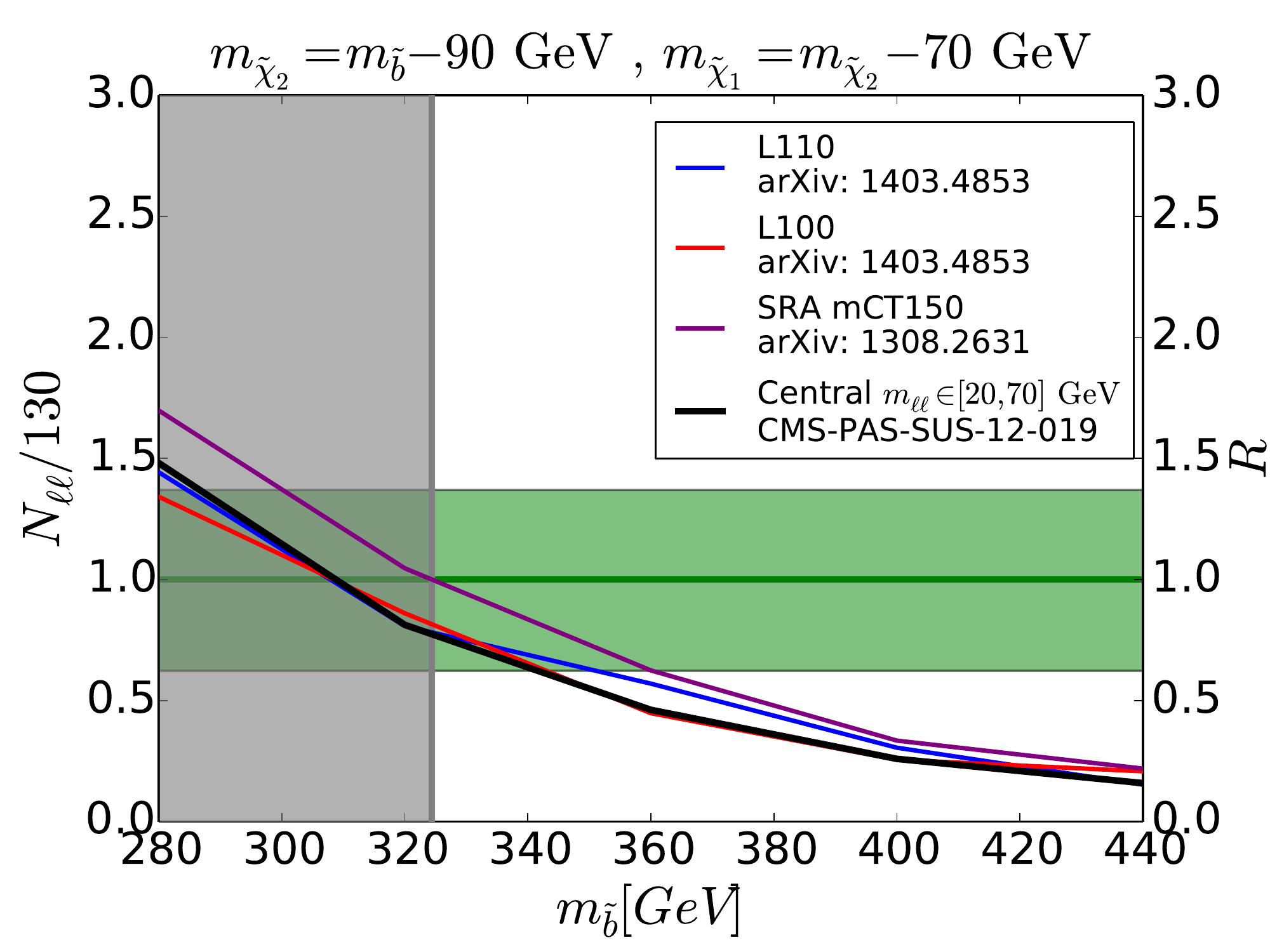}
\includegraphics[width=7.5cm]{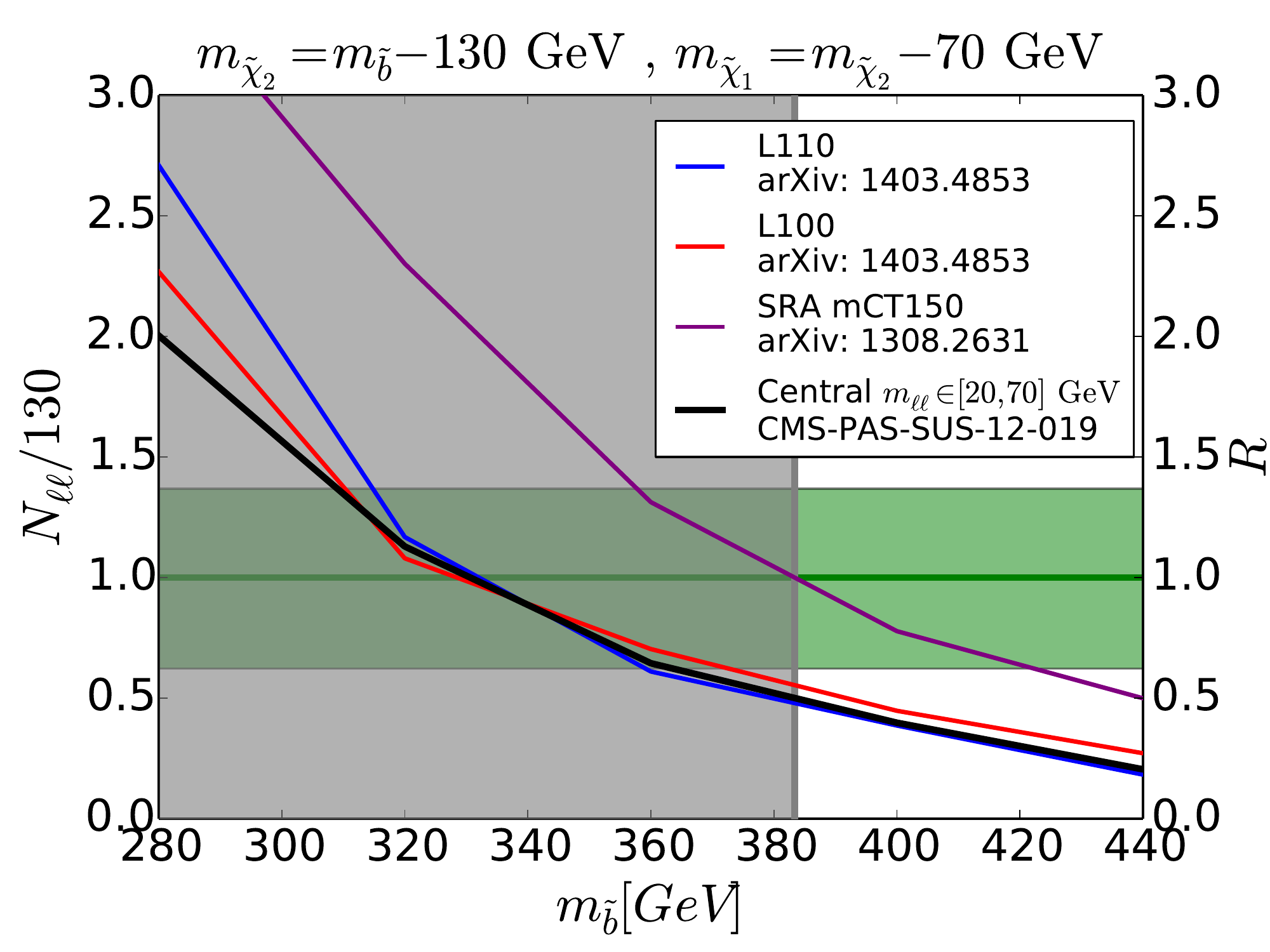}
\includegraphics[width=7.5cm]{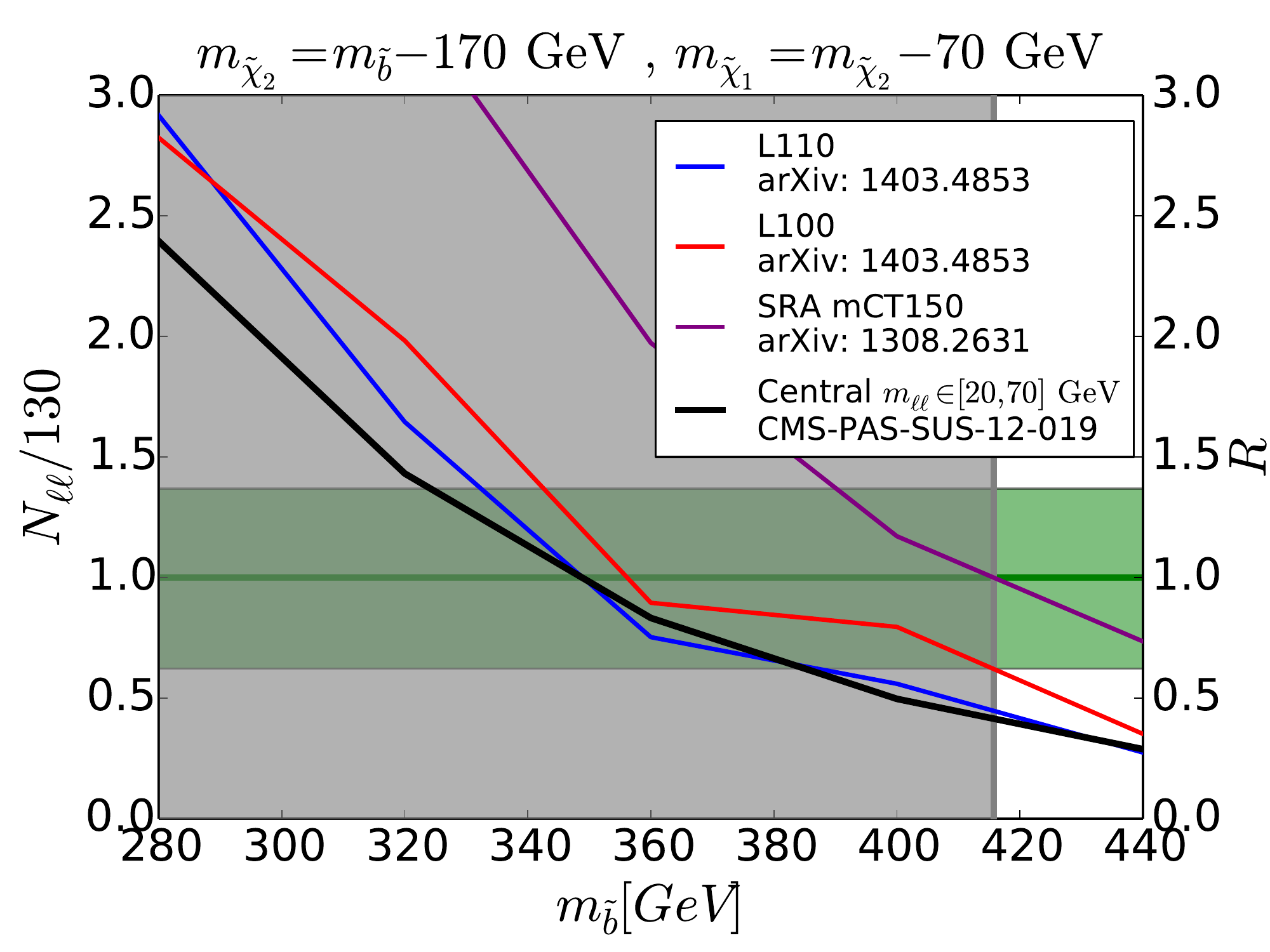}
 \caption{Signal rate and $R$-values for the off-shell $Z$-mediated right-handed sbottom-higgsino models. 
 } 
 \label{fig:r_sbottom_results}
\end{figure}

\begin{figure}[t]
 \centering
 \includegraphics[width=9cm]{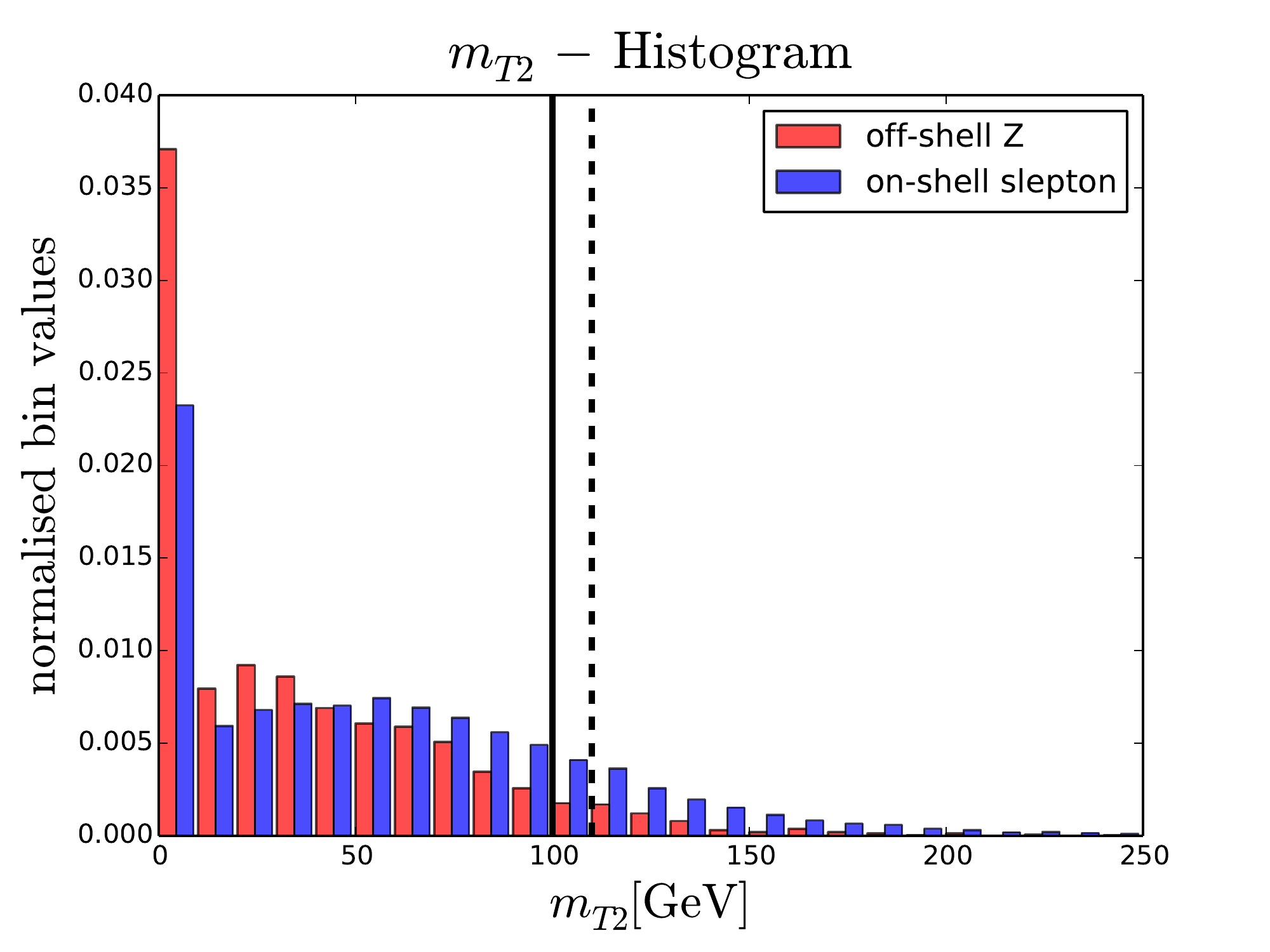}
 \caption{Histogram of $m_{T2}$-distribution for sbottom production for both the off-shell $Z$ and on-shell RH-slepton mediated case. The sbottom mass is 400 GeV and $m_{\ntwo} = 230$~GeV.  For the off-shell slepton mediated case we have $m_{\tilde{\mu}_R}=190$~GeV and $m_{\none}=151$~GeV. The black vertical line indicates the cut for the limiting signal region SRA mCT150.}
 \label{fig:mT2_histo}
\end{figure}

In Fig.~\ref{fig:r_sbottom_results} we show our results again in terms of $N_{\ell \ell}/130$ and $R$.
First we note that the strong constraint from L100 and L110 observed in the on-shell slepton models 
is relaxed.
To understand this we compare the distributions of $m_{T2}$, a kinematical variable used both in the L100 and L110 signal regions,
between the on-shell right-handed slepton (blue) and off-shell $Z$ models (red) in Fig.~\ref{fig:mT2_histo}
at similar mass spectra.
We take $(m_{\tilde b_1}, m_{\ntwo}) = (400, 230)$ GeV 
and fix $m_{\none}$ such that $m_{\rm edge} \simeq 78$ GeV 
for both models.  For the on-shell slepton model we take $m_{\tilde \ell} = 190$ GeV.
In Fig.~\ref{fig:mT2_histo} we see that the off-shell $Z$ model tends to give smaller 
$m_{T2}$ compared to the on-shell slepton model.
The solid (dashed) vertical black line represents the event selection cut on the $m_{T2}$ variable 
employed in the L100 (L110) signal region.
As can be seen, the off-shell $Z$ model is less sensitive to the the L100 and L110 signal regions
than the on-shell slepton model.  

What can also be seen from Fig.~\ref{fig:r_sbottom_results} is that for $\Delta m > 90$ GeV
the SRA mCT150 signal region in the ATLAS di-bottom analysis \cite{1308.2631} is constraining and 
most of the preferred region of the dilepton excess is indeed disfavoured by this signal region.  
This signal region looks for two energetic $b$-jets with $p_T > 130$ and $50$~GeV in events with $m_{\rm CT} > 150$ GeV\footnote{ 
$m_{\rm CT} \equiv \sqrt{ (E_T^{b_1} + E_T^{b_2})^2 - ({\bf p}_T^{b_1} - {\bf p}_T^{b_2})^2}$,
where $E_T$ and ${\bf p}_T$ are the transverse energy and the transverse momentum vector, respectively.}
 and $\met > 150$.
Events containing an electron ($p_T > 7$~GeV) or a muon ($p_T >  6$~GeV) are rejected in this analysis. 
This signal region is more constraining for larger $\Delta m$ because the event selection requires two energetic $b$-jets. 

For $\Delta m = 50$ and 90 GeV we find the regions where the observed excess can be explained at 1-$\sigma$ level
without $R > 1$ from other searches. 
This result is consistent with the findings reported in \cite{1410.4998}.
However these regions are already in tension with other searches.
In particular the ATLAS stop search \cite{1403.4853} and the ATLAS di-bottom search \cite{1308.2631} give $R \lsim 1$ in these regions.

\begin{figure}[t]
\centering
\includegraphics[width=7.5cm]{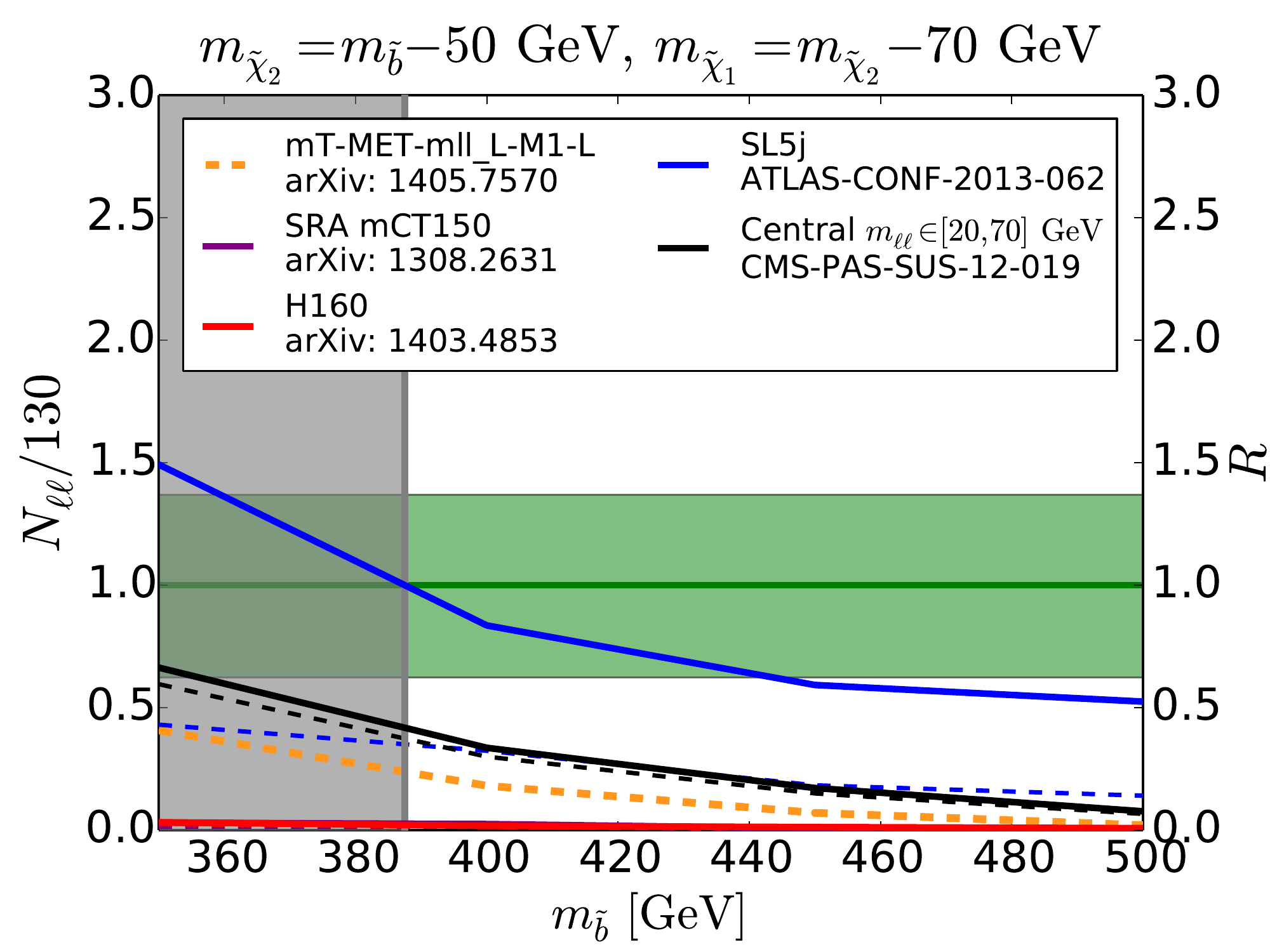}
\includegraphics[width=7.5cm]{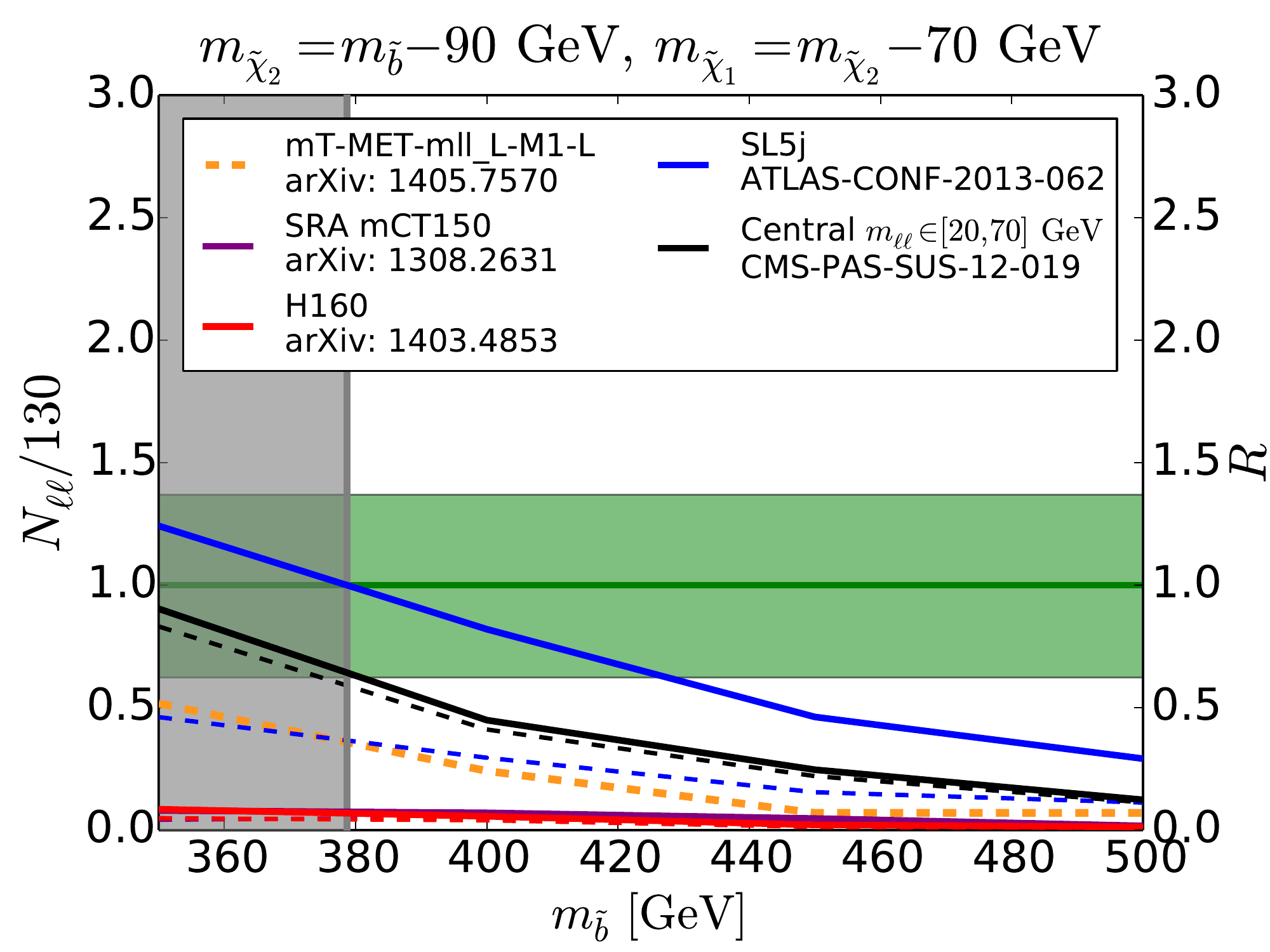}
\includegraphics[width=7.5cm]{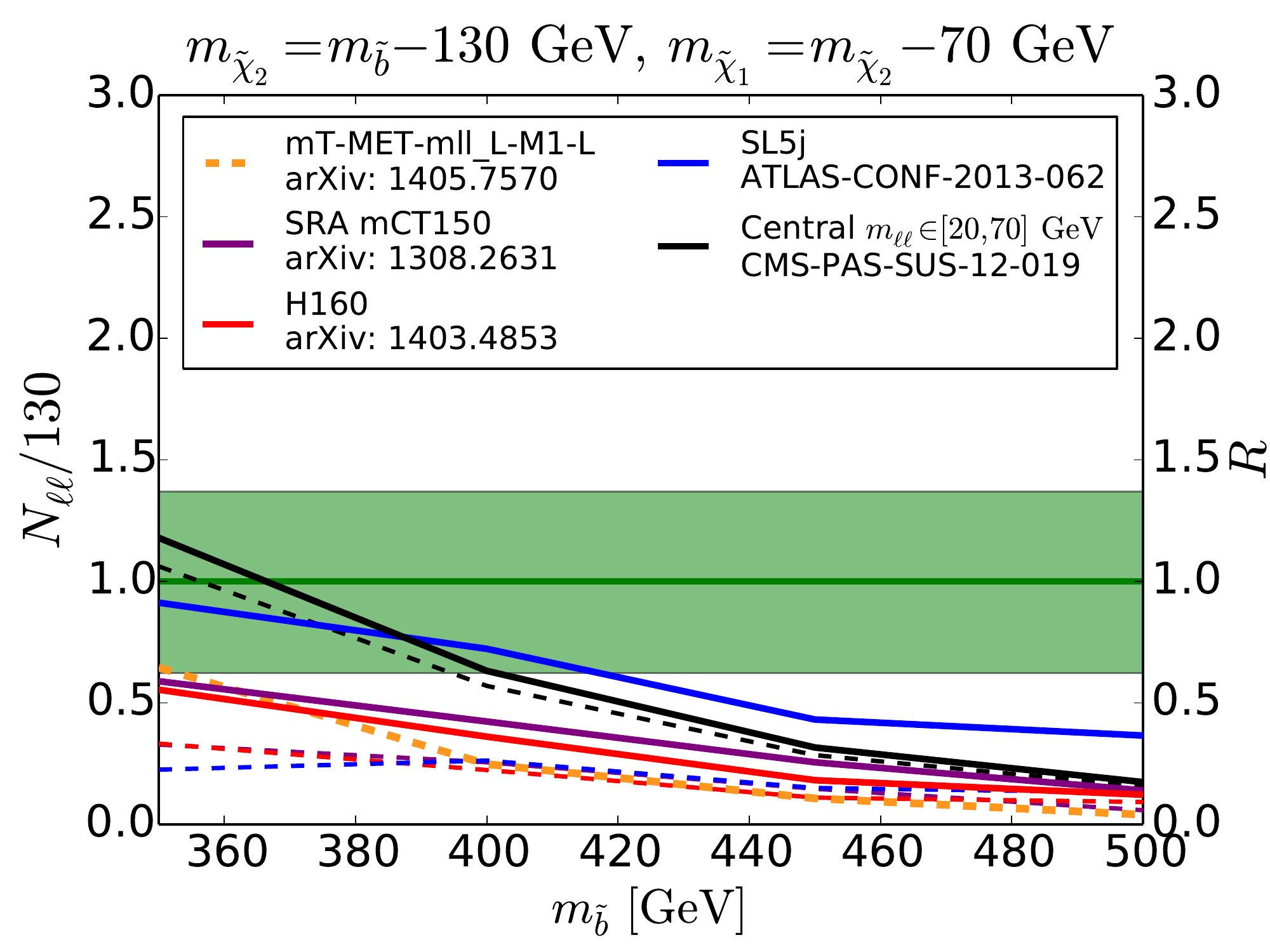}
\includegraphics[width=7.5cm]{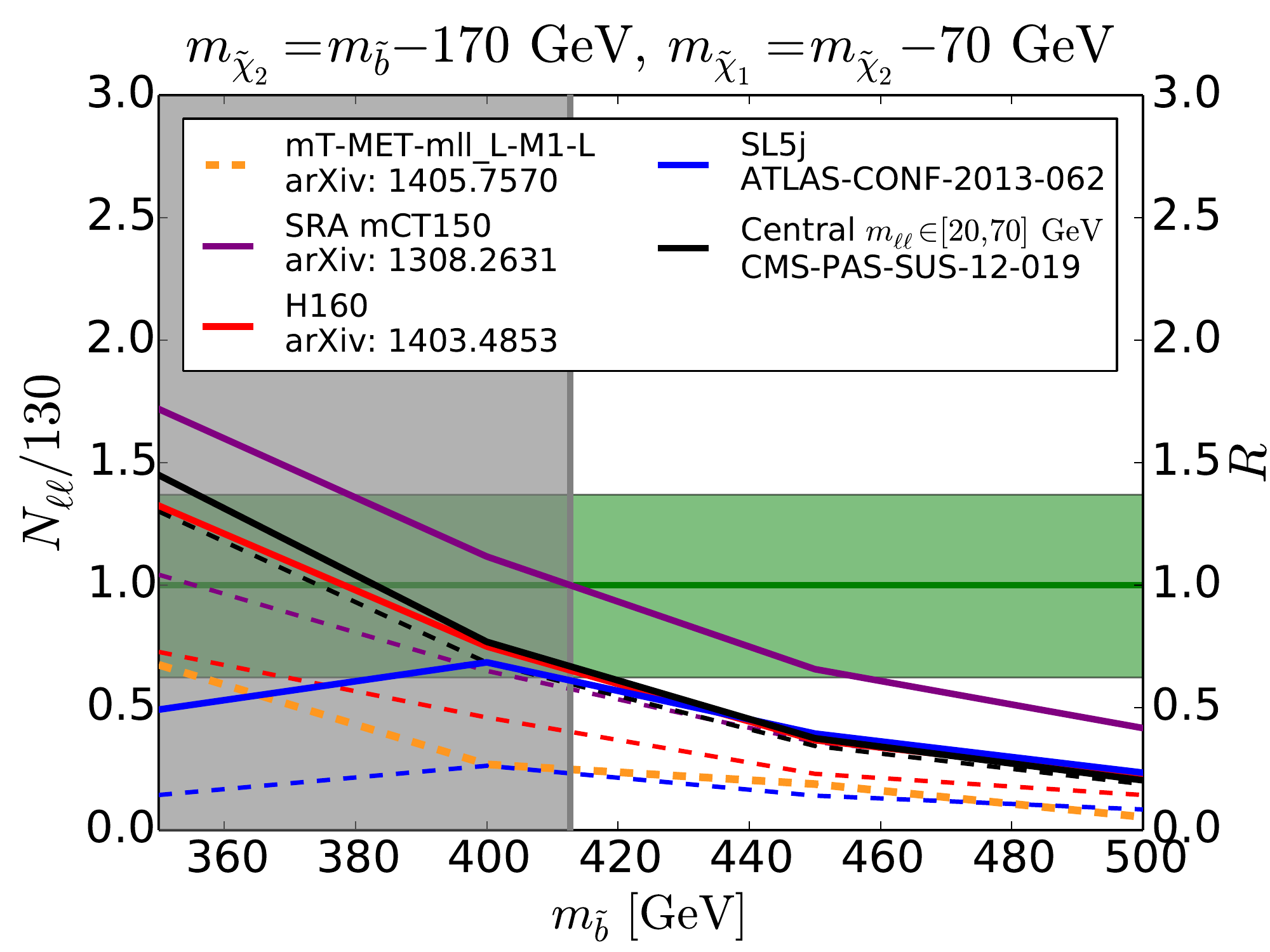}
 \caption{Signal rate and $R$-values for the off-shell $Z$-mediated left-handed sbottom model. Pure sbottom production is indicated by dashed lines and combined sbottom and stop production by solid lines.
 } 
 \label{fig:sbottom_Zresults}
\end{figure}

In Fig.~\ref{fig:sbottom_Zresults} we show $N_{\ell \ell}/130$ and $R$ as functions of $m_{\tilde b_1}$
in the left-handed sbottom model where $\ntwo$ ($\cone$) is assumed to be Wino-like and decays predominantly to an off-shell $Z$ ($W$) and a Bino-like $\none$. 
We again show the results for four different mass gaps
and fix $m_{\none} = m_{\ntwo} - 70$ GeV to fit the central value of the counting experiment. 
As we have mentioned in section \ref{sc:susy}, we assume the presence of the top squark, $\tilde t_1$, with $m_{\tilde t_1} = m_{\tilde b_1}$.
The solid curves in Fig.~\ref{fig:sbottom_Zresults} represent the results with both $\tilde b_1 \tilde b^*_1$ and $\tilde t_1 \tilde t^*_1$ production processes. To see the impact of the tilde $\tilde t_1 \tilde t^*_1$ production on the result, we also plot the contribution to $N_{\ell \ell}/130$ and $R$ 
from $\tilde b_1 \tilde b^*_1$ by dashed curves.

One can see from Fig.~\ref{fig:sbottom_Zresults} that for $\Delta m = 50$ and 90 GeV the model is strongly constrained by the SL5j signal region in the ATLAS jets plus 1-2 lepton analysis \cite{ATLAS-CONF-2013-062}. This signal region requires a soft single electron (muon) with $p_T \in [10, 25]$ $([6, 25])$ GeV and veto additional electron (muon) with $p_T > 10$ (6) GeV.  It also requires $\geq 5$ jets with $p_T > (180, 25, 25, 25, 25)$ GeV. 
The SL5j signal region is more sensitive to the $\tilde t_1 \tilde t^*_1$ topology where one of the stops decays hadronically $\tilde t_1 \to b \cone \to b W^* \none \to b q q' \none$ and the other decays leptonically $\tilde t_1 \to b \cone \to b W^* \none \to b \ell \nu \none$, because event selection requires a single lepton.
We also note that the SL5j signal region becomes less sensitive for larger $\Delta m$ 
because the leptons from the stop cascade decay chain are boosted in this case and do not pass the low $p_T$ requirement ($< 25$ GeV) efficiently.
However, for larger $\Delta m$ the SRA mCT150 signal region becomes constraining.
In particular the preferred region of the dilepton excess is disfavoured by this signal region at $\Delta m = 170$ GeV.   

As a result we find a good fit to the dilepton excess at $\Delta m = 130$ GeV and $m_{\tilde b_1} \in [350, 400]$ GeV,
although this region is already in tension with the SL5j signal region in the ATLAS jets plus 1-2 lepton analysis.
In addition, let us note that there is an additional constrain on the $\none-{\tilde t_1}$ mass plane 
from CMS single-lepton analysis \cite{1308.1586}, which is not included in our analysis.
This analysis does not use the cut-and-count method but rather uses a BDT multivariate method,
which prevents us from implementing this analysis.
While recasting this analysis is out of the scope of this work, it is worthwhile to deduce its constraint on our models. 
Specifically, the exclusion contour on the $\none-{\tilde t_1}$ mass plane with chargino mass fixed at $m_{\cone}=0.25~m_{\tilde t_1} + 0.75~m_{\none}$  in the CMS analysis is most relevant to the allowed parameter space in our study ($m_{\cone}\simeq 0.3~m_{\tilde t_1} + 0.7~m_{\none}$). 
At $m_{\tilde t_1} \simeq 380$ GeV the CMS analysis excludes $m_{\none} \lsim 200$ GeV,
whilts $m_{\none} = 180$ GeV at $m_{\tilde t_1} = 380$ GeV in the bottom left plot ($\Delta m = 130$ GeV) in Fig~\ref{fig:sbottom_Zresults}. 

\begin{figure}
\centering
\includegraphics[width=12cm]{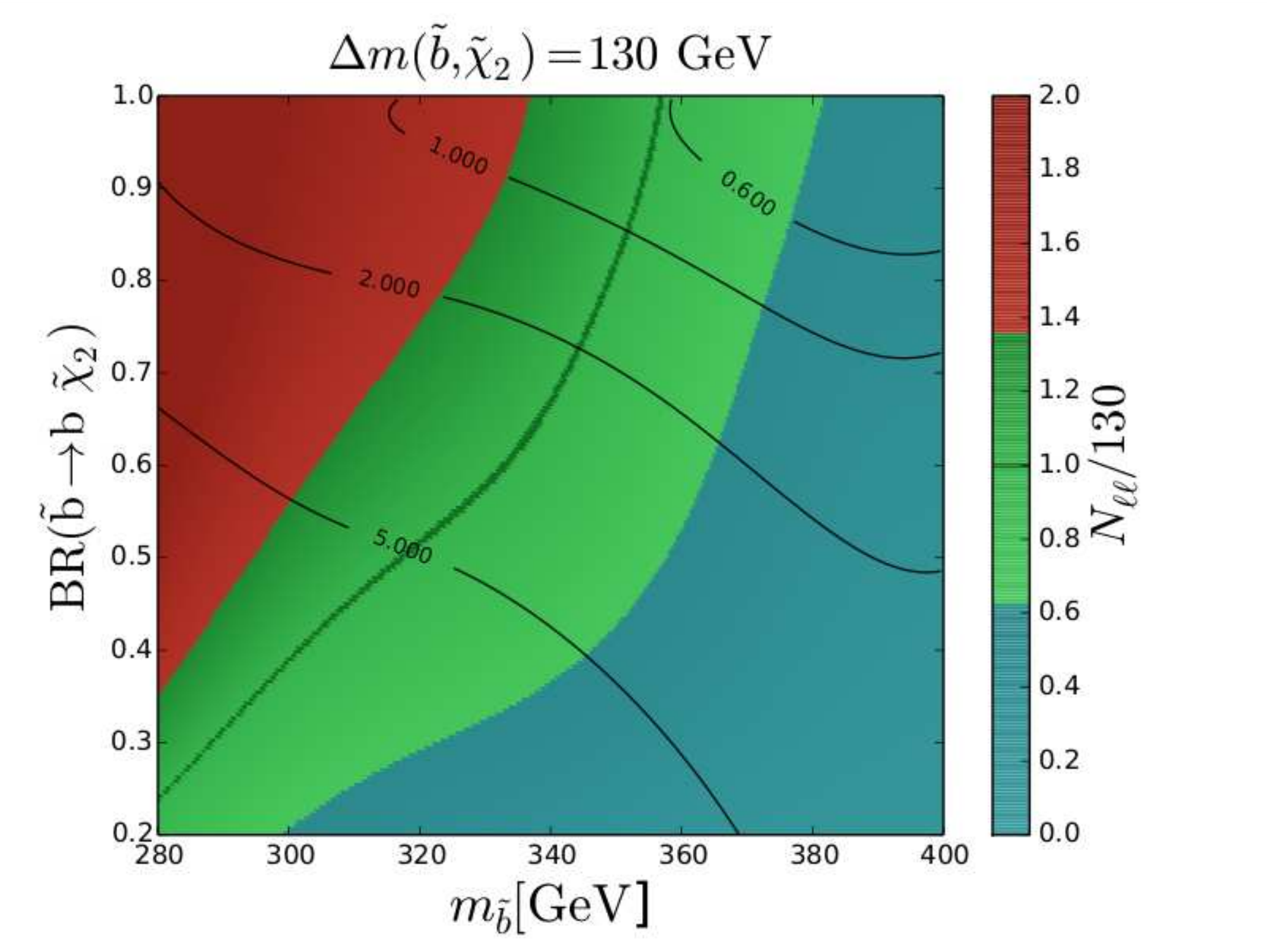}
 \caption{Variation of $\tilde{b}_1$ branching ratio into $\ntwo$ for off-shell $Z$ mediated left-handed sbottom scenario with no stop production. The color indicates $N_{\ell \ell}/130$ and the black curves are lines of constant $R_{\rm max}$. Only large wino branching ratios can provide a good fit to the excess.
 } 
 \label{fig:sbottom_2d}
\end{figure}

It is interesting to note that the difference between the black solid and black dashed curves are small, whereas the difference is large 
amongst the blue solid and blue dashed curves in the bottom left plot ($\Delta m = 130$ GeV) in Fig~\ref{fig:sbottom_Zresults}.
This means that the $\tilde b_1 \tilde b_1^*$ production gives the main contribution to the dilepton excess,
while the model is disfavoured mainly by the additional $\tilde t_1 \tilde t_1^*$ production.
Before concluding our study we show in Fig.~\ref{fig:sbottom_2d}
the contribution to the dilepton excess and the constraint from other searches
in the $m_{\tilde b_1}$ versus $BR(\tilde b_1 \to b \ntwo)$ plane
concerning only the $\tilde b_1 \tilde b_1^*$ production.
In this study we assume $BR(\tilde b_1 \to b \none) = 1 - BR(\tilde b_1 \to b \ntwo)$.
The region is divided into 3 colours, red, green and blue,
which correspond to under, good and over fit of the dilepton excess, respectively.  
The $R$ value of the most constraining signal region is shown in the black contours.
One can see that a good fit is found for $m_{\tilde b_1} \in [340, 380]$ GeV and
$BR(\tilde b \to b \ntwo) \gsim 0.8 $
without having $R > 1$ from other searches.
Within our exploration we did not find the models where the sbottom is mostly right-handed 
and  $BR(\tilde b \to b \ntwo) \gsim 0.8$.
However, this result indicates that
models that have a large cross section of the topology equivalent to 
$\tilde b_1 \to b \ntwo \to b Z^* \none$ 
can in principle explain the CMS dilepton excess avoiding constraints from other ATLAS and CMS direct SUSY searches.

\section{Conclusions}
\label{sc:con}

One straightforward supersymmetric interpretation of the observed dilepton excess by CMS~\cite{CMS-PAS-SUS-12-019,Khachatryan:2015lwa} is the cascade decays of light-flavour and bottom squarks.
In this paper, we studied and tested the viability of promising SUSY models by deriving constraints on these from various direct SUSY searches using the automated simulation tool {\tt Atom}.  

In order to obtain a contribution to the dilepton excess from SUSY events, we considered the decay of the second lightest neutralino, $\ntwo$, via either an off-shell $Z$ or an intermediate on-shell slepton. The $\ntwo$ itself arises from a light-flavour squark or sbottom decay. 
We investigated in total six possible simplified models, see figures~\ref{fig:squark_model},~\ref{fig:sbottom_slep} and~\ref{fig:sbottom_Z}.

We found that all of these models are already in strong tension with the experimental data once we demand a good fit to the dilepton excess. 
In particular strong limits arise from an earlier neglected ATLAS stop search~\cite{1403.4853} with identical final state topology. 
This analysis alone rules out the interpretation of the excess in terms of an intermediate (left- or right-handed) on-shell slepton for both light squark and sbottom production, see left panel of Fig.~\ref{fig:squark_results}, Fig.~\ref{fig:sbottom_Lslepton} and Fig.~\ref{fig:sbottom_Rslepton} respectively.
We showed that if multijet plus missing energy searches are taken into account, the off-shell $Z$ scenario with squark production is strongly disfavoured and noted that it is not able to give a sizeable contribution to the dilepton signal region, as can be seen in the right panel of Fig.~\ref{fig:squark_results}.

We confirmed the result reported in \cite{1410.4998} and showed that 
the right-handed sbottom model with Higgsino-like $\ntwo$ and $\nthree$ decaying predominantly into an off-shell $Z$
can explain the dilepton excess at 1-$\sigma$ level, 
although the model is already in tension with the ATLAS di-bottom search and the ATLAS stop search, as can be seen in Fig.~\ref{fig:r_sbottom_results}.
This tension can be ameliorated if the left-handed sbottom model with Wino-like $\ntwo$ is considered.
However, the left-handed stop is necessarily introduced in this model
and that creates another tension with the ATLAS jets plus 1-2 lepton analysis as be seen in Fig.~\ref{fig:sbottom_Zresults}.

We also showed in Fig.~\ref{fig:sbottom_2d} that in a simplified model that only has sbottom production
the dilepton excess can be explained avoiding constraints from other searches
in the region where $m_{\tilde b_1} - m_{\ntwo} \sim 130$ GeV, $m_{\tilde{b}} \sim 360$~GeV and $BR(\tilde b_1 \to b \ntwo) \gsim 0.8$,
although we did not find a corresponding model point in the context of the MSSM within our exploration.
This results may indicate that a more non-trivial SUSY scenario should be considered to explain
the CMS dilepton excess.

{\bf Note added:}
Shortly after this paper was submitted to arXiv, CMS updated 
their result \cite{Khachatryan:2015lwa} and reported most of the excessive events 
are observed associated with at least one $b$-jet. 
This new information further disfavours the squark scenario, which does not change our conclusion.
Shortly after the CMS update, ATLAS released their new analysis of the jets plus SFOS dilepton channel \cite{Aad:2015wqa}.
They explicitly looked at the signal region employed in the CMS dilepton analysis and did not find any significant excess.
This casts a doubt that observed dilepton excess is merely due to the statistical fluctuation or background mismodeling.
The next run of the LHC will provide a definitive answer to this question.

\appendix

\section{Validation} \label{ap:validation}

Here we show the validation results of our implementation of CMS-PAS-SUS-12-019/1502.06031~\cite{CMS-PAS-SUS-12-019,Khachatryan:2015lwa} and the ATLAS stop search with two lepton final state~\cite{1403.4853}.

The benchmark point considered in the CMS analysis has a sbottom of mass 400 GeV decaying via $\tilde{b} \rightarrow\ntwo b$ with 100\%. The second lightest neutralino then undergoes an off-shell $Z$ decay with SM branching ratios. We show the good agreement between the CMS results and our implemented analysis in {\tt Atom} in table~\ref{tbl:val_cms}. There, we give the event numbers in the central and forward signal regions as quoted by the CMS collaboration and their ratio to our results obtained with {\tt Atom}.

Additionally, we provide validation results for the stop search because of the strong constraints that we derive from this analysis. The ATLAS benchmark scenario consists of a stop decaying to $\cone + b$ with 100\% probability followed by a decay of $\cone$ via a $W$ into $\none$ and Standard Model particles. We show our validation in table~\ref{tbl:val_atlas}. In this table we present event numbers for the same-flavour (SF) and different-flavour (DF) case as given by ATLAS and their ratio to our results in the column {\tt Atom}/Exp.

\begin{table}[h] 
\centering 
\begin{tabular}{|l|l|l|l|l|}
\hline
 $( m_{\tilde{b}},m_{\ntwo} ) = ( 400,150 )~\GEV$ & Central &{\tt Atom}/Exp   & Forward &  {\tt Atom}/Exp
\\
\hline
$N_{\rm jets}~\geq 2$(no $\met$ requirement)& $242.7\pm 2.8$ & 1.04 & $34.2\pm 1.1$& 0.77
\\
$N_{\rm jets}~\geq 3$(no $\met$ requirement)& $186.2\pm 2.5$ & 1.09 & $25.6\pm 0.9$&  0.76
\\
\hline
$\met > 100~\GEV$(no $N_{\rm jets}$ requirement)& $152.5\pm 2.1$ & 1.03 & $19.8\pm 0.8$& 0.98
\\
$\met > 150~\GEV$(no $N_{\rm jets}$ requirement)& $85.0\pm 1.5$ & 0.93 & $10.4\pm 0.5$& 0.87
\\ 
\hline
Signal region & $132.4 \pm 2.0$& 1.031 &$17.0 \pm 0.7$ & 0.937
\\
\hline
\end{tabular} 
\caption{ Validation table for our implementation of the CMS-PAS-SUS-12-019/1502.06031 analysis~\cite{CMS-PAS-SUS-12-019,Khachatryan:2015lwa} in {\tt Atom}.
\label{tbl:val_cms}
}
\end{table}

\begin{table}[h] 

\begin{tabular}{|l|l|l|l|l|}
\hline
 $( m_{\tilde{t}},m_{\cone},m_{\none} ) = ( 400, 390, 195 )~\GEV$ & SF & {\tt Atom}/Exp   & DF &  {\tt Atom}/Exp
\\
\hline
$\Delta \phi > 1$ & 1834.9 & 1.09 & 2390.1 & 1.06
\\
$\Delta \phi_{b}$ & 1402.8 & 1.07 & 1800.5 &  1.07
\\
\hline
$m_{T2} > 90~\GEV$ & 396.5 & 1.02 & 500.0 & 1.09
\\
$m_{T2} > 120~\GEV$ & 211.8 & 1.01 & 284.4 & 1.1
\\ 
\hline
$m_{T2} > 100~\GEV$, $p_{T,{\rm jet}} > 100~\GEV$ & 21.7 & 1.4 & 35.0 & 0.99
\\
$m_{T2} > 110~\GEV$, $p_{T,{\rm jet}} > 20~\GEV$ & 86.0 & 0.95 & 116.1 & 0.89
\\
\hline
\end{tabular} 
\caption{ Validation table for our implementation of the ATLAS stop search with two leptons~\cite{1403.4853} in {\tt Atom}.
\label{tbl:val_atlas}
}
\end{table} 

\vspace{2cm}

\section{Parameter Values for pMSSM scan}
\label{ap:pmssm}

In table~\ref{tbl:pmssm} we give additinal pMSSM input parameters as well as the sum of the branching ratio of $\tilde{b}_1$ to $\ntwo$, $\nthree$. These points were used to scan the right-handed sbottom-Higgsino model. Calculation of the physical SUSY masses and branching ratios was done using SPheno~\cite{Porod2003,Porod2011}. 

\begin{table}[h]
 \centering 
\begin{tabular}{|l|l|l|l|l|}
\hline
    & $\Delta m (\tilde{b}, \ntwo)=$50 & $\Delta m (\tilde{b}, \ntwo)=$90 & $\Delta m (\tilde{b}, \ntwo)=$130 & $\Delta m (\tilde{b}, \ntwo)=$170 \\
 \hline
 $m_{\tilde{b}}=$280 &(229,157,0.48) &(183,157,0.63) &(140,157,0.69) &(96,157,0.63) \\
\hline
 $m_{\tilde{b}}=$320 &(270,215,0.48) &(228,215,0.63) &(183,215,0.69) &(140,215,0.67) \\
\hline
 $m_{\tilde{b}}=$360 &(307,263,0.46) &(269,263,0.63) &(228,263,0.69) &(183,263,0.7) \\
\hline
 $m_{\tilde{b}}=$400 &(345,310,0.46) &(307,310,0.63) &(269,310,0.7)  &(228,310,0.71) \\
\hline
\end{tabular}
\caption{Additional information about the right-handed sbottom-higgsino model. We give values for ($\mu$, $m_{\tilde{b}_R}$,$\sum_{i=2,3} BR(\tilde{b} \rightarrow b \, N_i)$) for each model point. All masses and mass differences are given in GeV. See text for more details. \label{tbl:pmssm}}
\end{table}

\vspace{1cm}

\section*{Acknowledgment}
We are grateful to J.~S.~Kim, K.~Rolbiecki, and J.~Tattersall for collaborations during the early stages of this work. P.G. is supported by an ERC grant. S.P.L. is supported by JSPS Research Fellowships for Young Scientists and the Program for Leading Graduate Schools, MEXT, Japan. 
The work of K.S. was supported in part by the London Centre for Terauniverse Studies (LCTS), using funding from the European Research Council via the Advanced Investigator Grant 267352.

\end{document}